\documentclass[nobibnotes,amssymb,aps,twocolumn,superscriptaddress,floatfix,longbibliography]{revtex4-2}
\usepackage[dvips]{graphicx}
\usepackage{dcolumn}
\usepackage{bm}
\usepackage{amsmath}
\usepackage{amssymb}
\usepackage{ulem}
\usepackage{xcolor}
\usepackage{float}
\graphicspath{{./figs/}}
\usepackage{empheq}
\usepackage{lipsum}

\usepackage[compact]{titlesec}
\titlespacing{\section}{10pt}{*2}{*2}
\titlespacing{\subsection}{10pt}{*2}{*2}
\titlespacing{\subsubsection}{10pt}{*4}{*4}

\newcounter{boxlblcounter}  

\begin{document}
\title{Optical Interference Effect in Strong-field Electronic Coherence Spectroscopy}

\author{Eleanor Weckwerth}
\address{Stanford PULSE Institute, SLAC National Accelerator Laboratory, Menlo Park, California 94025, USA}
\address{Department of Physics, Stanford University,
Stanford, California 94305, USA}
\author{Andrew J. Howard}
\address{Stanford PULSE Institute, SLAC National Accelerator Laboratory, Menlo Park, California 94025, USA}
\address{Department of Applied Physics, Stanford University,
Stanford, California 94305, USA}
\author{Chuan Cheng}
\address{Stanford PULSE Institute, SLAC National Accelerator Laboratory, Menlo Park, California 94025, USA}
\address{Department of Physics, Stanford University,
Stanford, California 94305, USA}
\author{Ian Gabalski}
\address{Stanford PULSE Institute, SLAC National Accelerator Laboratory, Menlo Park, California 94025, USA}
\address{Department of Applied Physics, Stanford University,
Stanford, California 94305, USA}
\author{Aaron M. Ghrist}
\address{Stanford PULSE Institute, SLAC National Accelerator Laboratory, Menlo Park, California 94025, USA}
\address{Department of Applied Physics, Stanford University,
Stanford, California 94305, USA}
\author{Salma A. Mohideen}
\address{Stanford PULSE Institute, SLAC National Accelerator Laboratory, Menlo Park, California 94025, USA}
\address{Department of Physics, Stanford University,
Stanford, California 94305, USA}
\author{Chii-Dong Lin}
\address{J. R. Macdonald Laboratory, Department of Physics, Kansas State University, Manhattan, Kansas 66506, USA}
\author{Chi-Hong Yuen}
\thanks{Corresponding author: cyuen2@kennesaw.edu}
\address{J. R. Macdonald Laboratory, Department of Physics, Kansas State University, Manhattan, Kansas 66506, USA}
\address{Department of Physics, Kennesaw State University, Marietta, Georgia, 30060, USA}
\author{Philip H. Bucksbaum}
\thanks{Corresponding author: phbuck@stanford.edu}
\address{Stanford PULSE Institute, SLAC National Accelerator Laboratory, Menlo Park, California 94025, USA}
\address{Department of Physics, Stanford University,
Stanford, California 94305, USA}
\address{Department of Applied Physics, Stanford University,
Stanford, California 94305, USA}

\date{\today}

\begin{abstract}
We have investigated strong-field-induced electronic coherences in argon and molecular nitrogen ions created by high-intensity, few-cycle infrared laser pulses.
This is a step toward the long-sought goal of strong-field coherent control in molecular chemistry.
We employed high-intensity, few-cycle infrared laser pulses in a pump-probe setup to investigate
a recent prediction that 
electronic coherences in nitrogen molecules change the ion yields vs. pump-probe delay.
[Yuen and Lin, Phys. Rev. A 109, L011101 (2024)]. The predicted coherence signals in molecular nitrogen could not be resolved above the optical interference of the pump and probe pulses; a simultaneous measurement clearly resolved the induced cation fine-structure coherence in strong-field-ionized argon. The results of our comparison with simulations suggest that optical interference effects manifest differently in each ionic species and must be carefully accounted for when interpreting experimental data.
We found that nonsequential double ionization in the low-intensity region of the focal volume can reduce the visibility of coherence generated by two-pulse sequential ionization, and we quantify the importance of pulse shape and spectral characteristics for isolating the desired coherence signals.
\end{abstract}

\maketitle

\section{Introduction~\label{sec1}}

The characteristics and time evolution of electronic coherence in molecules has been a bottleneck in the pursuit of strong-field ultrafast control of chemical reactions~\cite{Calegari2023_Open, Ivanov2021_Concluding}.
As electronic coherence drives electron motion in molecules, manipulating electronic coherence is crucial for controlling ultrafast chemistry.
The survival of electronic coherence against nuclear motion leads to charge migration, while its loss results in charge transfer~\cite{Cederbaum1999_Ultrafast, Remacle2006_electronic, Arnold2017_Electronic}.
Therefore, mastering control of electronic coherence could revolutionize our ability to influence chemical reactivity and functionality in molecules~\cite{Remacle1998_Charge, Lepine2014_Attosecond, Pianowski2022_Molecular}, shape processes in quantum biology such as vision~\cite{Schoenlein1991_first, Polli2010_Conical}, light harvesting~\cite{Herek2002_Quantum, Wang2019_Quantum, Cao2020_Quantum}, avian magnetoreception~\cite{Lambert2013_Quantum}, and find applications in quantum sensing~\cite{Wang2022_Atomicscale} and quantum computing~\cite{Wasielewski2020_Exploiting}.
Strong-field control is especially desirable because its non-perturbative nature generates a stronger molecular response, leading to more effective control.
Within a single strong-field pulse, both single- and multi-photon interactions emerge that couple a plethora of electronic states and generate electronic coherence.
A fundamental approach to controlling electronic coherence involves using the laser field to drive electric dipole coupling between two electronic states.
Therefore, achieving strong-field control requires diagnostic measurements to probe the electronic coherence between two laser-coupled electronic states.

Numerous experiments have been conducted to initiate and observe charge migration and transfer, but only a few schemes feature a mechanism that can probe the electronic coherence between laser-coupled states under the influence of nuclear motion.
High-harmonic spectroscopy was utilized to observe and control such electronic coherence within one optical cycle of an infrared (IR) laser pulse~\cite{Smirnova2009_High, Kraus2015_Measurement, He2022_Filming, He2023_Attosecond}, but the effects of nuclear motion, which typically set in a few femtoseconds (fs), could not be resolved.
Other experiments employed an isolated attosecond extreme ultraviolet (XUV) pulse pump and intense IR probe~\cite{Calegari2014_Ultrafast, Lara-Astiaso2018_Attosecond, Herve2021_Ultrafast}, attosecond XUV pulse train pump-probe~\cite{Okino2015_Direct, Fukahori2020_Ultrafast}, or a femtosecond x-ray pulse pump-probe~\cite{Barillot2021_Correlationdriven, Schwickert2022_Electronic}, where a superposition of electronic states is created by ionizing the molecules.
Due to the broad bandwidth of the attosecond pulses, several electronic states are typically populated. 
This significantly complicates the interpretation of coherence signals, making the probing mechanisms difficult to understand. 
A better-understood experimental approach for probing charge migration and transfer is attosecond transient absorption spectroscopy (ATAS)~\cite{Kobayashi2020_Attosecond, Kobayashi2020_Coherent, Matselyukh2022_Decoherence}, where the intense IR pump creates a superposition of electronic states in the neutral or ionic species and the attosecond XUV or soft x-ray probes the electronic coherence by absorption.
However, later results showed that ATAS can probe electronic coherence between dipole-forbidden states~\cite{Golubev2021_Corevalence, Kobayashi2022_Theoretical, Yuen2024_Rotation}, so that the signal from coherence between laser-coupled states is likely to be suppressed.
Another recent experiment employed shaped octave-spanning ultrafast laser pulses to create a superposition of neutral excited states and probe their electronic coherence through multiphoton absorption and ionization~\cite{Kaufman2023_Longlived}.
It was shown that electronic coherence between states with energy separations of one photon can survive the nuclear motion.
While the multiphoton pumped states are susceptible to further strong-field control, their results hint that a strong-field pump-probe scheme could probe electronic coherence between states of different parity.

A recent theoretical development by Yuen and Lin~\cite{Yuen2024_Probing} further motivates the use of strong-field pump-probe schemes to initiate and observe charge migration and charge transfer.
They calculated that a few-cycle intense IR pump pulse can create a partially coherent superposition of ionic states through multiorbital tunnel ionization~\cite{Yuen2023_Coherence}.
A more intense IR probe pulse can then deplete the remaining neutral population, drive the laser couplings between ionic states, and trigger further tunnel ionization to form dissociative dications.
The variation in electronic coherence with pump-probe delays alters the strength of their laser couplings, thereby changing the transient population of excited ionic states that ionize to form dissociative dications.
Thus the electronic coherence between laser-coupled states is imprinted onto the observables from dissociative dications, such as kinetic energy release or branching ratios, which may be readily observed using few-cycle intense IR pulse pairs and coincidence ion detection.

We measured the yield of the N$^{+}$/N$^{+}$ channel with a pair of intense few-cycle IR pulses at variable pump-probe delays, as recommended by the strong-field dissociative sequential double ionization (DSDI) spectroscopy proposed by Yuen and Lin~\cite{Yuen2024_Probing}. We found that the vibronic beating of N$_{2}^{+}$ they predicted cannot be identified in the Fourier spectrum of the measured yield. The N$^{+}$/N$^{+}$ yield instead closely resembles the measured joint second harmonic generation signal in a BBO crystal.
To diagnose the measured results, we measured the yield of Ar$^{2+}$ using the same setup and compared with additional theoretical models to determine the primary effects present in both signals.
While the visibility of electronic coherence signals can be obscured by both optical interference and the details of the strong-field physics of the molecular species, our comparisons indicate that the optical effects are much more relevant to the visibility of the electronic coherence signal.

By combining our experimental results with theoretical simulations, we conclude that optical interference between the pedestal of one pulse and the main pulse of another significantly contaminates the electronic coherence signal from N$_{2}^{+}$. 
We also determine that background subtraction of the optical interference effect in N$_{2}^{+}$ from Ar$^{+}$ is not feasible, and that nonsequential double ionization (NSDI) from the probe pulse additionally contaminates the signal.
Our findings emphasize that to achieve strong-field DSDI spectroscopy for electronic coherence in molecules, it is essential to clean up the pedestal of few-cycle IR pulses and control the focal volume between the pump and probe pulses to prevent NSDI.

This article is arranged as follows.
Section~\ref{sec2} presents the background about the molecular and atomic targets, and Section~\ref{sec3}  provides details on experimental and theoretical approaches.
Our main results are discussed in Section~\ref{sec4}.
Section~\ref{sec5} discusses the possibility of background subtraction of optical interference effects. 
Section~\ref{sec6} addresses the role of non-sequential double ionization.
Section~\ref{sec7} examines the optical interference effects in ideal DSDI spectroscopy using cleaned Gaussian pulses.
Finally, we present our conclusions in Section~\ref{sec8}.

\section{Background~\label{sec2}}

\begin{figure}[t]
	\centering
	\includegraphics[width=0.8\linewidth]{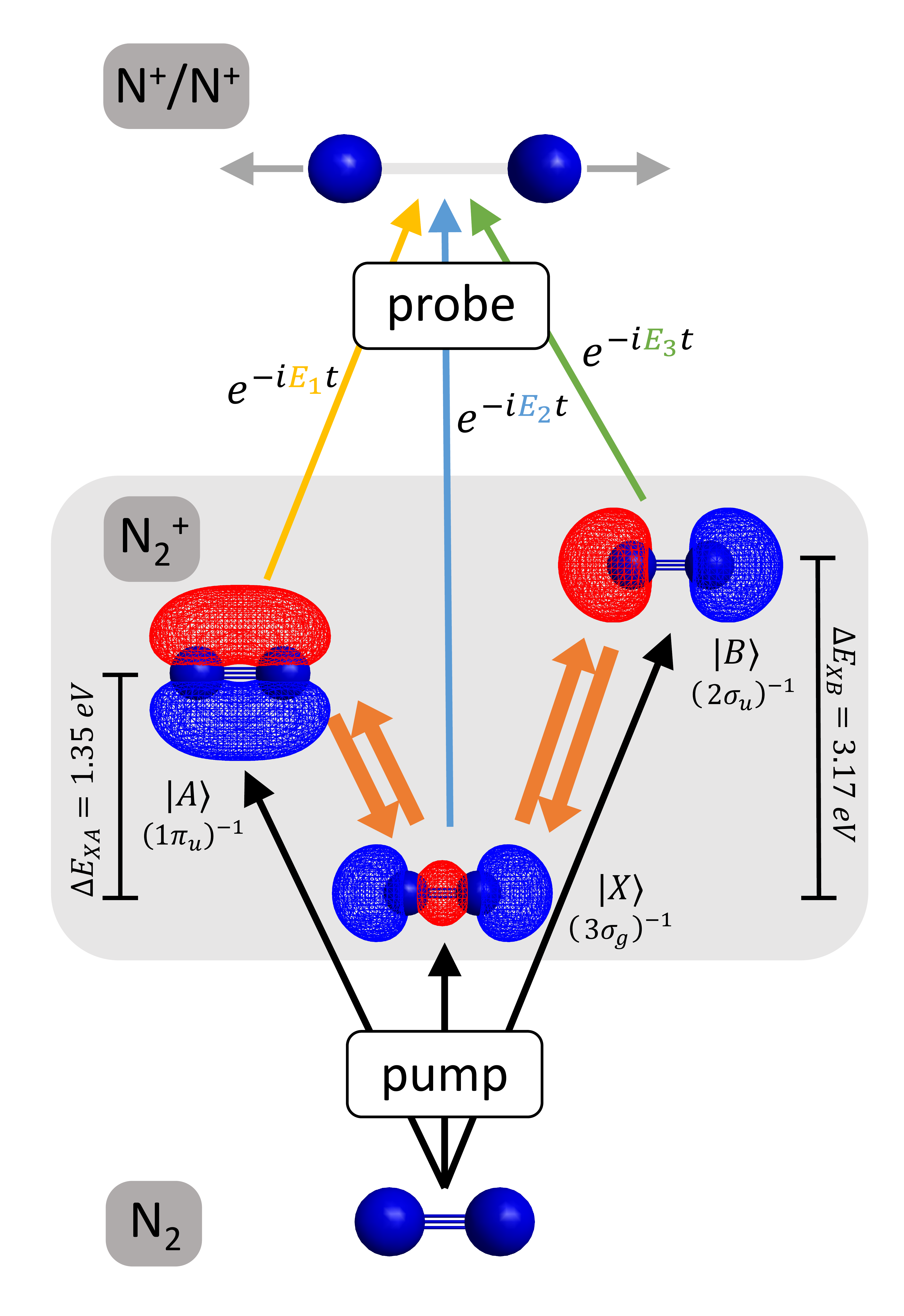}
	\caption{ Pump/probe scheme for the experiment. The pump pulse drives the molecule to one of several cationic states, followed by the probe pulse at a variable delay, which induces second ionization to produce the final dication. The probe pulse additionally couples the ionic states of opposite symmetry, shown with orange arrows. The energy separation of the coupled states are shown on the vertical black bars. Evidence of coherent oscillations at these energies present in the experimental data would be a signature of electronic coherence.}
	\label{diagram_andexp}
\end{figure}

The observation of electron motion in N$_{2}^{+}$ has been reported by He \textit{et al.} using high-harmonic spectroscopy~\cite{He2022_Filming} and Okino \textit{et al.} using attosecond pulse train pump-probe~\cite{Okino2015_Direct}.
Here, we describe only the relevant N$_{2}^{+}$ states in the strong-field DSDI spectroscopy.
A schematic diagram illustrating the pump-probe scheme for N$_{2}$ is shown in Figure~\ref{diagram_andexp}.
Details about the electronic structure of N$_{2}^{+}$ and N$_{2}^{2+}$ can be found in Refs.~\cite{Yuen2022_Densitymatrix, Jia2024_Improved} and references therein.
The pump pulse singly ionizes the N$_{2}$ molecule (ionization potential $I_{p} =$ 15.6 eV), producing N$_{2}^{+}$ in the low-lying electronic states, $X^{2}\Sigma_{g}^{+}$, $A^{2}\Pi_{u}^{+},$ and $B^{2}\Sigma_{u}^{+}$ states.
For simplicity, hereafter they are denoted as the $X, A$, and $B$ states, respectively.

When in their ground vibrational states, the $X$ and $A$ states and the $X$ and $B$ states are separated by 1.1 eV and 3.2 eV, respectively. There is an additional strongly Franck-Condon allowed transition between the $\nu=0$ of $X$ and the $\nu=1$ of $A$ with a 1.3 eV energy separation.
The laser couples the $X-A$ states and $X-B$ states with transition dipole moments of 0.25$\hat{x}$~\cite{Langhoff1987_Theoretical} and 0.75$\hat{z}$~\cite{Langhoff1988_Theoretical} atomic units (a.u.).
After the pump pulse, their nuclear wavepacket $|\chi_{i} (t) \rangle$ ($i=X, A$, or $B$) on their respective potential energy curves lead to a varying electronic coherence $\rho_{ij} (t) \propto \langle \chi_{j} (t) |\chi_{i} (t) \rangle$, which contains the $X-A$ and $X-B$ vibronic beating. 
The probe pulse drives the laser coupling between the $X-A$ and $X-B$ states and triggers the second ionization, producing a dissociative dication measured as a coincident pair of N$^{+}$ ions.
Ideally, the $X-A$ vibronic beating around 1.1 and 1.3 eV will be dominant in the Fourier spectrum of the N$^{+}$/N$^{+}$ yield, while the $X-B$ beating around 3.2 eV will be smaller in amplitude but still measurable~\cite{Yuen2024_Probing}. In the fixed-nuclei approximation, only the 1.3 eV $X-A$ transition corresponding to the vertical energy separation at the neutral equilibrium distance is considered.

A similar few-cycle intense IR pump-probe experiment has been reported in Ar~\cite{Fechner2014_StrongFielda}.
The ionization potential for Ar is about 15.8 eV.
Upon ionizing the $m=0$ orbital by the pump pulse, the two lowest spin-orbit states are coherently populated, where the energy separation between $P_{1/2}$ and $P_{3/2}$ states is about 0.18 eV.
The spin-orbit wave packet leads to a periodic oscillation in the population of the $m=|1|$ hole state, which is preferentially ionized by a more intense probe pulse ($I_p$ = 27.7 eV).
The previous experiments indeed observed the spin-orbit wave packet dynamics, but they also observed the optical interference effect between the pump and probe pulses.
Therefore, we expected that the Fourier spectrum of the Ar$^{2+}$ yield to be dominated by the 0.18 eV peak and the peaks from the laser frequency spectrum.
Similar optical interference effects could be present in the yield of N$^{+}$/N$^{+}$.


\section{Methods~\label{sec3}}
\subsection{Experiment~\label{sec3.1}}
The experiment employed 40-fs 800-nm 1-kHz Ti:sapphire laser pulses which were spectrally broadened in a 2.5-m long, 500-$\mu$m diameter stretched hollow core fiber differentially pumped with gaseous Ar to a maximum pressure of 4 psig~\cite{Robinson2006_generation}. The broadened pulses were then re-compressed to 6.5-fs duration using chirped mirrors. The beam was split using a Mach-Zehnder interferometer to produce two identical pulses at variable interpulse delays between 30-115 fs. The spectral interference between the two pulses was measured after the interferometer to extract the delays with high precision. One arm of the interferometer contained additional pellicle ND filters to reduce the power of the pump pulse by a factor of 3.1 without adding additional dispersion.

The pulse duration was characterized using a second-harmonic generation (SHG) dispersion scan with BK7 wedges~\cite{Miranda2012_Simultaneous,Sytcevich2021_Characterizing}. This method allows us to reconstruct the temporal intensity profile $I(t)$ of the pulse, which can be used in simulations to replicate the experimental conditions. An additional pulse characterization technique used was an interferometric intensity autocorrelation, which measures the intensity of second-harmonic generation (SHG) produced as a function of the interpulse delay. Pulse characterization and retrieval is discussed in more detail in Appendix~\ref{app:pulse}.

The pulse pairs were focused by a $f=5$ cm concave mirror to peak intensities of $3.70 \times 10^{14}$ W/cm$^{2}$ and $1.15 \times 10^{15}$ W/cm$^{2}$ for the pump and the probe pulses, respectively. The focusing mirror was located inside a vacuum chamber containing a dilute gaseous sample of a nearly equal mixture of N$_{2}$ and Ar ($1.3 \times 10^{-9}$ Torr), with the partial pressure of each gas scaled independently to produce a similar count rate of both N$_{2}^{+}$ and Ar$^{+}$ ions. The ionic fragments produced were accelerated by charged plates held at a high voltage towards a detector consisting of a triple-stack of microchannel plates and a Roentdek delay-line hex-anode~\cite{Jagutzki2002_Multiple}. The overall ion count rate was held at an average of 1.2 ions/shot to ensure true coincidences among detected ions. The breakup momentum of detected ions is calculated using the measured position and time-of-flight data for each event.

The 3D momentum of each ion is used to filter coincidences by momentum conservation. The kinetic energy of those coincident fragments can also then be used to calculate the Kinetic Energy Release (KER) of each breakup event, which is used to determine the dication state of the molecule before dissociation.

\subsection{Theory~\label{sec3.2}}

The entire strong field pump-probe process for N$_2$ and Ar was simulated using the density matrix approach for sequential double ionization (DM-SDI).
Details about the theory can be found in Refs.~\cite{Yuen2024_Probing, Yuen2024_Theory}.
The essence of the DM-SDI model is that ionized electrons are neglected, leaving the residual ions to be open systems.
This is motivated by the fact that SDI is driven by a laser rather than by the rescattering of an ionized electron.
To further simplify calculations, nuclei of the molecule are frozen in the presence of external fields.
This approximation is valid when the pulse duration is much shorter than the vibrational period of the molecule, which is about 13 fs for N$_2$.
To model the intense laser interaction with the target, we numerically solve the von Neumann-Liouville equation for the density matrices $\rho^{(q)}$, where $q=0,1$, or 2 for neutral, ion, or dication.
The equations of motion are
\begin{align}
    \frac{d}{dt} \rho^{(q)}(t) = -\frac{i}{\hbar} [H^{(q)}(t),\rho^{(q)}(t)] + \Gamma^{(q)}(t).
    \label{Eq:EOM}
\end{align}
The density matrices are represented in terms of electronic states, in which $\rho^{(q)}$ for different charge states are block-diagonal.
The Hamiltonian $H^{(q)}$ consists of the field-free term and the electric dipole coupling $-\Vec{d}^{(q)} \cdot \Vec{E}$, where $\Vec{E}$ is the laser field.
The ionization matrices $\Gamma^{(q)}$, which take different forms for N$_2$ and Ar, account for the population and coherence changes due to tunnel ionization.
After solving Eq.~\eqref{Eq:EOM}, the population and coherence in the electronic states for charge $q$, which are diagonal and off-diagonal elements of $\rho^{(q)}$, are obtained.

The simulations employed two types of laser fields, either the experimentally retrieved pulses or Gaussian pulses. When Gaussian pulses were used, the peak intensities followed the values given in the Experiment section, with a central wavelength of 750 nm and a pulse duration of 4.5 fs unless stated otherwise. The text specifies the laser field used for each simulation discussed in the following sections.

\subsubsection{DM-SDI model for N$_2$}
For the pump process of N$_2$, Eq.~\eqref{Eq:EOM} is solved with the initial conditions $\rho^{(0)}(t_0) = 1$ and $\rho^{(1)}(t_0) = \rho^{(2)}(t_0)= 0$.
Only the ground state is considered for the neutral while the $X, A$, and $B$ states are considered for the ion.
Details about the dication states can be found in Refs.~\cite{Yuen2022_Densitymatrix, Jia2024_Improved}.
The ionization matrices for N$_2$ are based on the molecular Ammosov-Delone-Kraniov theory~\cite{Tong2002_Theory} and the weak field asymptotic theory~\cite{Tolstikhin2011_Theory}.
Their forms can be found in Refs.~\cite{Yuen2023_Coherence, Yuen2024_Theory}.
Under the Franck-Condon approximation and the fixed-nuclei approximation, the nuclear wave packets for the ionic states after the end of the pump pulse at $t=t_1$ are $|\chi_0 \rangle$, the $v=0$ vibrational wave function of neutral N$_2$. The vibronic states are populated according to their Franck-Condon factors with $|\chi_0 \rangle$.
The nuclear wave packet $|\chi_i (t)\rangle$ is evolved in time analytically~\cite{Yuen2024_Probing}.
For $t>t_1$, the ionic density matrix is propagated according to $\rho^{(1)}_{ij}(t) = \rho^{(1)}_{ij}(t_1) \langle \chi_j(t)|\chi_i(t) \rangle$~\cite{Yuen2024_Probing}.

The probe process starting from $t=t_2$ is modeled by solving Eq.~\eqref{Eq:EOM} with the initial conditions $\rho^{(0)}(t_2) = \rho^{(0)}(t_1)$, $\rho^{(1)}(t_2)$, and $\rho^{(2)}(t_2) = 0$~\cite{Yuen2024_Probing}.
The obtained dication yields are mapped to the yield of N$^+$ + N$^+$ and the kinetic energy release~\cite{Yuen2022_Densitymatrix, Jia2024_Improved}.
In the simulations, the start and end time of the pump or probe pulse is taken to be $\mp \tau$ with respect to the peak of the pulse, where $\tau$ is the pulse duration.

When the optical interference effect between the pump and probe pulse is considered, the above approach cannot be rigorously employed as the overlap of the pulses could trigger tunnel ionization and laser couplings.
For such a case, we study the effect by freezing the nuclei of N$_2$ at all times.
The pump-probe process will be simulated in a single run by solving Eq.~\eqref{Eq:EOM} with the initial conditions $\rho^{(0)}(t_0) = 1$ and $\rho^{(1)}(t_0) = \rho^{(2)}(t_0)= 0$, and the field $\Vec{E}$ will consist of the pump and probe pulses.

\subsubsection{DM-SDI model for Ar}
A comprehensive density matrix approach for SDI of noble gas atoms was reported in Ref.~\cite{AdrianN.Pfeiffer2013_Calculation}.
Here, we further simplify the approach by considering Ar$^+$ is formed by ionizing the electron with $m_s = -1/2$ from the $m_l=0$ orbital.
Spin-orbit coupling leads to the formation of $P_{j=3/2, m=1/2}$ and $P_{j=1/2, m=1/2}$ states for Ar$^+$, where $m_l$ is restricted to 0 and 1.
Therefore, we represent the $ls$ basis by the $m_l=0$ and 1 hole states and the $jm$ basis by the $j=1/2$ and 3/2 states.
They are related by
\begin{align}
    \begin{pmatrix}
    |j= 3/2 \rangle \\
    |j= 1/2 \rangle
    \end{pmatrix}
=
M
    \begin{pmatrix}
    |m_l= 0 \rangle \\
    |m_l= 1 \rangle
    \end{pmatrix},
\end{align}
where
\begin{align}
M = \frac{1}{\sqrt{3}}
    \begin{pmatrix}
    \sqrt{2} & 1 \\
    -1 & \sqrt{2}
    \end{pmatrix}.
\end{align}
The final Ar$^{2+}$ state is represented only by the $3p_0^{-1} 3p_{1}^{-1}$ hole state.

The entire pump-probe process of Ar is simulated in a single run with the initial conditions $\rho^{(0)}(t_0) = 1$ and $\rho^{(1)}(t_0) = 0$.
Eq.~\eqref{Eq:EOM} is solved in the $jm$ basis, where the Hamiltonian matrix is diagonal.
The ionization matrix for the ion in the $jm$ basis is
\begin{align}
    \Gamma^{(jm)}(t) = M \Gamma^{(ls)}(t) M^T
\end{align}
with
\begin{align}
    \Gamma^{(ls)}(t) &= \rho^{(0)}(t) \delta_{i0} W_0 (t) - \rho^{(ls)}_{11}(t) \delta_{i1} W_{1} (t) \label{eq:argon1} \\
     \rho^{(ls)}(t) &= M^T \rho^{(jm)}(t) M.
\end{align}
In the above, $W_0$ and $W_1$ is the tunnel ionization rate of $m_l=0$ orbital from Ar and $m_l=1$ hole state of Ar$^+$, with the over barrier suppression~\cite{Tong2005_Empirical}.
Upon solving Eq.~\eqref{Eq:EOM}, the yield of Ar$^{2+}$ at the last time step is obtained by
\begin{align}
    \rho^{(2)}(t_f) = 1 - \rho^{(0)}(t_f) - \rho^{(ls)}_{00}(t_f) - \rho^{(ls)}_{11}(t_f).
    \label{eq:argon2}
\end{align}

\section{Main Results~\label{sec4}}

\begin{figure}[htb]
	\centering
	\includegraphics[width=\linewidth]{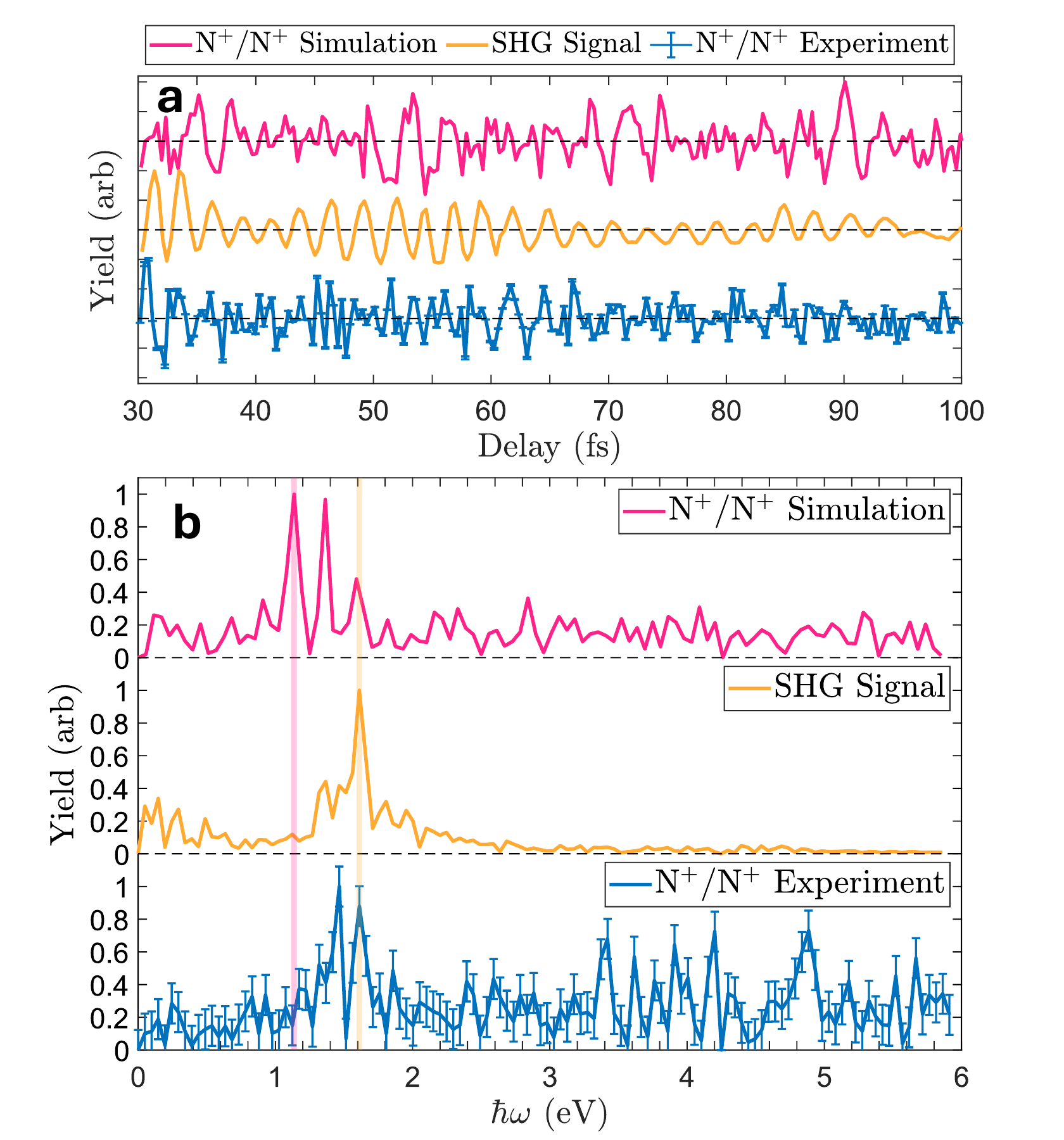}
	\caption{ Comparison of experimental N$^{+}$/N$^{+}$ ion coincidence channel (blue) with expected result from DM-SDI simulation~\cite{Yuen2024_Probing} (pink) and measured second-harmonic generation (SHG) signal from the experiment (orange). \textbf{(a)} Yield as a function of interpulse delay for all three signals. \textbf{(b)} Results of a Fast-Fourier Transform applied to the data in (a). Vertical lines designate the maximum frequency peak in the simulation and SHG signal. In both panels, the systematic experimental noise at 4.25 eV is filtered out as described in Appendix~\ref{app:stage}. Error bars on the experimental data reflect statistical uncertainties.
    }
	\label{fft_NpNp_shg}
\end{figure}

The electronic coherences between the $X$, $A$, and $B$ states of the N$_{2}^{+}$ cation are probed by measuring the yield of coincident N$^{+}$ ions as a function of pump-probe delay. The measured yield of N$^{+}$ for each measured delay in the experiment is shown as the blue curve in Fig.~\ref{fft_NpNp_shg}a. The yield is filtered to only include ions detected in coincidence from an N$^{+}$/N$^{+}$ breakup event, as described in the Experiment section. Some features become clear in the delay-dependent data. For example, the yield is maximized near the beginning of the delay range, where the two pulses are most overlapped in time, indicating that optical interference (OI) between the two pulses is likely present in the signal. Additionally, there is a significant high frequency contribution present across all delays, but for particular delay ranges, such as 60-65 fs, the dominant frequency appears to be one with a period of about 2.7 fs corresponding to the central frequency of the Ti:sapph pulses, again indicating the presence of OI.

To better understand the measured data, we compare to two other relevant delay-resolved yields. The predicted signal based on DM-SDI calculations by Yuen and Lin~\cite{Yuen2024_Probing} is shown in pink in Fig.~\ref{fft_NpNp_shg}a.
This simulation was performed using Gaussian pulses with a random carrier envelope phase (CEP), which is fixed within the pulse pair for each delay.
The OI effect was neglected so that variation of electronic coherence in N$_{2}^{+}$ due to nuclear motion is accounted for.
Around 70 fs, the phase of the N$^+$ yield clearly does not match the experiment.
The maximum yield is around 90 fs rather than around 30 fs, as in the experiment.
Note that the randomized CEP introduces a noisy background as high as 40\% of the peak signal.
This background was absent in Ref.~\cite{Yuen2024_Probing}, in which the CEP was set to zero, and it covers the weak FFT signal around 3.2 eV from the $X-B$ coherence.

A third curve (orange) in Fig.~\ref{fft_NpNp_shg}a displays the joint second harmonic generation (SHG) intensity as a function of interpulse delay measured during the experiment.  For delays larger than 60 fs, peaks of the SHG signal are in phase with those of the experimental N$^+$ yield. The intensity of the SHG signal depends on the square of the fundamental pulse intensity, so it is nonlinearly affected by the temporal overlap of the pulses. Since it is measured with the same pulses and delays used in the experiment, it is a good representation of the OI effect that might be present in the yield of N$^+$.

We perform a Fast-Fourier Transform (FFT) to extract the oscillation frequencies present in the yields, which are shown in Fig.~\ref{fft_NpNp_shg}b. The uncertainties in the experimental N$^{+}$/N$^{+}$ ion yield and FFT spectrum, as well as in all other experimental data presented in this work, are determined from Poisson counting statistics combined with standard error propagation. A systematic experimental noise signal caused by the translation stage motion at about 4.25 eV is present and was filtered out of the frequency spectrum, discussed in more detail in Appendix~\ref{app:stage}. As suggested by the N$^{+}$/N$^{+}$ yield in panel (a), the FFT spectrum of the experimental N$^{+}$/N$^{+}$ channel includes a broad peak spanning 1.3 to 1.7 eV. However, it is not immediately clear whether OI is solely responsible for these peaks since there are expected electronic coherence signals in the N$^{+}$/N$^{+}$ channel within this range. Distinguishing between OI and coherence signals is a challenge that will be a focus of the discussion here.

The predicted signal shown in pink in Fig.~\ref{fft_NpNp_shg}b is based on an ideal case where no optical interference is present in the measured delay range. The two primary peaks in the frequency spectrum are the expected vibronic beatings at 1.1 and 1.3 eV, corresponding to the $X-A$ beating of the N$^{+}$ cation.
In contrast, the SHG signal shown in orange is entirely optical and is therefore shifted to a higher frequency around 1.55 eV, corresponding to the central laser wavelength. The primary frequency here is also visible in the time domain data as a constant frequency oscillation present across all delays. Neither the SHG signal nor the simulation alone reproduce the measured N$^{+}$/N$^{+}$ yield. There are no peaks above the noise floor at the same frequencies as the vibronic beatings in the simulated signal. There is also a larger contribution from high frequencies in the ion signal, some of which reach a similar level as the peaks in the OI region. In general, the primary N$^{+}$/N$^{+}$ yield is shifted more towards the frequencies of the SHG signal than towards the simulated frequencies, indicating that the measured signal is more consistent with the OI effect than with the $X-A$ beating.

We further resolved the kinetic energy release of the N$^{+}$/N$^{+}$ ions at different interpulse delays.
We found that OI effects persist in all channels but manifest differently.
A detailed comparison and discussion can be found in Appendix~\ref{app:KER}.

\section{Background Subtraction of Optical Interference Signal~\label{sec5}}

If the coherence signals are present but obscured by OI, one potential solution is to use the ion yield of another species that have similar ionization potentials $I_p$ and experiences the same fields to subtract out the OI contribution to the signal. One might expect N$_{2}$ and Ar to respond very similarly to OI effects since they have comparable $I_p$ : 15.6 eV for N$_{2}$ and 15.8 eV for Ar, and 25 to 30 eV for N$_{2}^{+}$ and 27.6 eV  for Ar$^{+}$. One would hope that the OI effects in the N$^{+}$/N$^{+}$ channel could be removed by an appropriate scaling of the Ar$^{2+}$ channel.

\begin{figure}[htb]
	\centering
	\includegraphics[width=\linewidth] {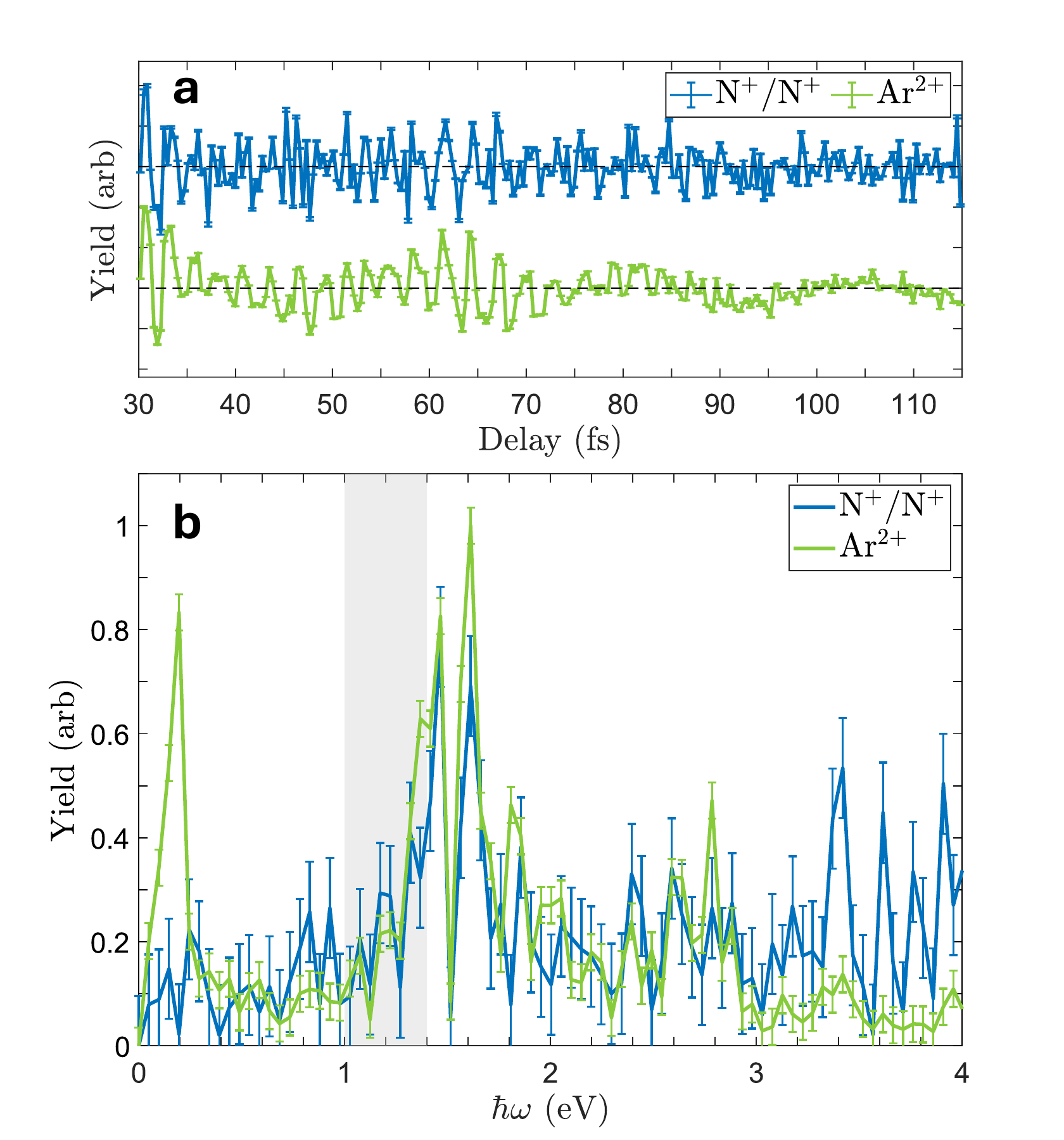}
	\caption{Comparison of N$^{+}$/N$^{+}$ coincidence channel and Ar$^{2+}$ channel time-resolved yields. \textbf{(a)} Ion yield versus pump-probe delay for N$^{+}$/N$^{+}$ ions and and Ar$^{2+}$ ions. \textbf{(b)} FFT amplitude of the data in (a). The SOWP oscillation in Ar$^{+}$ appears as a slow oscillation (period of approximately 20 fs) in the Ar$^{2+}$ delay data and as a 0.2 eV peak in the Fourier Transform. The shaded region from 1.0-1.4 eV is where the expected $X-A$ electronic coherence in the N$_{2}^{+}$ cation should appear. The FFT signals are normalized to the sum over the region from 2.2-3.0 eV, which has no expected vibronic or electronic levels in either species. Error bars on the experimental data reflect statistical uncertainties.}
	\label{NpNp_Arpp_exp}
\end{figure}

However, comparison of the measured ion yield for the two species shown in Fig.~\ref{NpNp_Arpp_exp} suggests that subtraction of the OI signal is not possible. The two species were measured simultaneously as described in Section~\ref{sec3.1}. Figure~\ref{NpNp_Arpp_exp}a contains the measured ion yield as a function of delay for both the N$^{+}$/N$^{+}$ coincidence channel and the Ar$^{2+}$ channel, which demonstrates a slow oscillation consistent with the spin-orbit splitting in the Ar$^{+}$ cation. Figure~\ref{NpNp_Arpp_exp}b shows the FFT amplitude of the data, where this spin-orbit wave packet (SOWP) feature appears as a prominent peak near 0.2 eV. The frequency spectrum shows other differences between the two species, including larger high-frequency contributions in the N$^{+}$/N$^{+}$ channel.

The noise level in both species should be similar because the total yield of both species matches to within a few percent. The region between 2.2-3.0 eV is not consistent with any vibronic or electronic splitting in either species, and is outside the expected OI region for the 800 nm pulses; thus, this region is expected to be zero in both species. The FFT amplitudes are normalized to the sum over this region to maintain a similar level of noise between the two species. The shaded region between 1.0-1.4 eV is where the expected $X-A$ electronic coherence for the N$_{2}^{+}$ cation would appear if it were detected. Since there are no clear peaks above the noise level in this region that are unique to N$^{+}$/N$^{+}$, we would have to rely on subtraction of the two channels to isolate any potential coherence signals. However, the energy range with the largest discrepancy between the two signals is in the OI range near 1.6 eV, which precludes our ability to use the Ar$^{+}$ curve above 0.2 eV as a measurement of solely OI. Clearly, there is a large difference in the OI effect for the two species that were expected to have similar tunnel ionization dynamics.

To understand the OI effects in N$_{2}$ and Ar, we first simulate the pump-probe scheme for Ar using the DM-SDI model as described in Sec.~\ref{sec3.2}. In the simulation, we used the laser pulse reconstructed from the experiment, with a random CEP fixed within the pulse pair for each delay.
We calibrate the model by adjusting the peak intensity of the pump pulse to match the vertical offset of the experimental Ar$^{+}$/Ar$^{2+}$ yield at different delays.
The simulated and experimental signal was first fitted to a sinusoid, which accounts for the SOWP oscillation present in the Ar$^{2+}$ channel.
The vertical offset of the fit then provides an intensity-related metric for the calibration.
We found that for peak intensities of about $3.70 \times 10^{14}$ W/cm$^{2}$ and $1.15 \times 10^{15}$ W/cm$^{2}$ for the pump and probe pulses, the vertical offset of the simulated signal agrees well with the experiment.

\begin{figure}[htb]
	\centering
	\includegraphics[width=\linewidth]{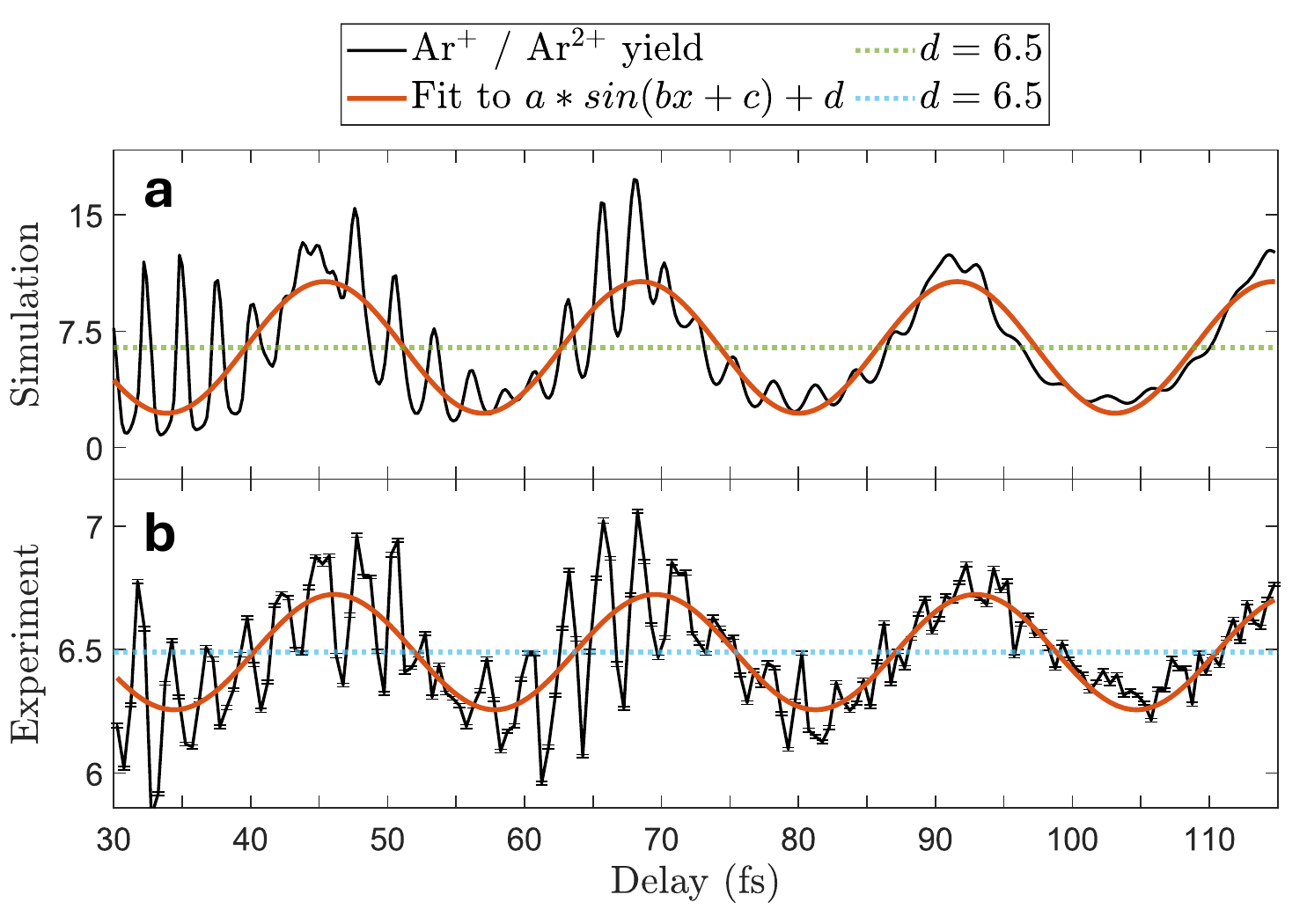}
	\caption{ Yield ratio of Ar$^{+}$/Ar$^{2+}$ channels for both \textbf{(a)} DM-SDI calculations and \textbf{(b)} experimental results. The y-axis ranges and scales of the two panels are different to emphasize the overall trends of the data. Error bars on the experimental data reflect statistical uncertainties.}
	\label{sim_exp_compare}
\end{figure}

The measured ratio of the Ar$^{+}$/Ar$^{2+}$ yields from both experimental data and simulated DM-SDI data is shown in Figure~\ref{sim_exp_compare}. 
We see that the DM-SDI model reproduces the qualitative behavior of the experimental results very well. The fitted oscillation periods of $23.1 \pm 0.4$ fs for the simulation and $23.4 \pm 0.5$ fs for the experiment are consistent with the expected SOWP frequency from the Ar$^{2+}$ channel.  In particular, the OI effects relative to the slower oscillation matches very well for the delays larger than 50 fs.

However, the depth of modulation of the SOWP oscillations differs significantly between the simulation and experiment.
The amplitude of the fits for the simulated data and experimental are $4.2 \pm 0.2$ and $0.23 \pm 0.03$, respectively.
This discrepancy strongly suggests that nonsequential double ionization (NSDI) by the probe pulse also contributes to the Ar$^{2+}$ signal, as this process is not sensitive to the SOWP dynamics.
Influence of NSDI on the pump-probe signal will be discussed in details in Sec.~\ref{sec6}.
Since the DM-SDI model neglects the contribution from NSDI, the pump and probe intensities may be overestimated in the calibration.

The excellent qualitative agreement between the model and experiment on the OI effects allows us to perform different simulations to understand why background subtraction of the OI signal is not feasible.
In addition to Ar and N$_2$, we also simulate the identical strong-field pump-probe scheme on Kr atom since it shares a similar SOWP dynamics as Ar.
The $I_p$ of Kr is 14.2 eV, which is lower than Ar. The energy separation of the two spin-orbit states $P_{3/2}$ and $P_{1/2}$ of Kr$^+$ is about 0.67 eV. The $I_p$ of Kr$^+$ is 24.8 eV, which is also lower than Ar$^+$.
On the other hand, as mentioned in Sec.~\ref{sec3.2}, to include the OI effects in the DM-SDI model, nuclei of N$_2$ must remain fixed.
As a result, only electronic beating of $X-A$ (1.3 eV), $A-B$ (1.8 eV), and $X-B$ (3.2 eV) could be presented in the FFT spectrum.
Accurately modeling the laser pulse in the simulation was prioritized because the experimental results shown in Fig.~\ref{fft_NpNp_shg} suggest that OI effects play a larger role in obscuring the coherence signals than nuclear motion effects.
All of the simulations were performed using the same parameters as in Fig.~\ref{sim_exp_compare}, with a reconstructed pulse to match the experimental conditions and the same pump and probe intensities.
The simulated signal for N$_2$ was averaged over molecular orientations.

\begin{figure}[htb]
	\centering
	\includegraphics[width=\linewidth]{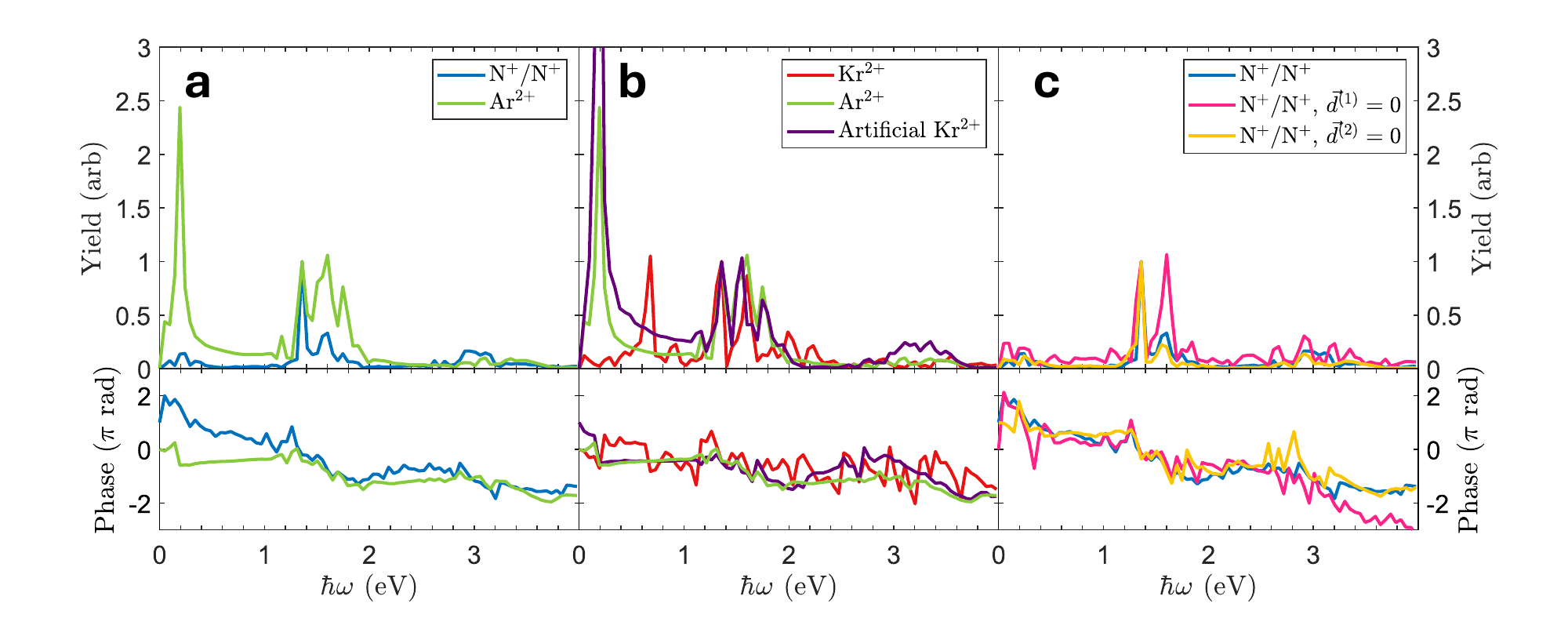}
	\caption{
	Fourier Transform amplitude and phase for DM-SDI simulated ion yields using experimental conditions, relative to the 1.3 eV peak. \textbf{(a)} N$^{+}$/N$^{+}$ ions in coincidence and Ar$^{2+}$ ions under the same conditions they were measured experimentally. \textbf{(b)} Kr$^{2+}$ ions and Ar$^{2+}$ ions with a third species that has the first and second $I_p$ of Kr but the spin-orbit energy separation of Ar. \textbf{(c)} N$^{+}$/N$^{+}$ ions in coincidence compared to the same species with laser couplings in one charge state turned off.}
	\label{oi_sim_compare}
\end{figure}

Figure~\ref{oi_sim_compare}a compares the calculated FFT signal for simulated yields of N$^{+}$/N$^{+}$ ions in coincidence and Ar$^{2+}$ ions. The peak at approximately 0.2 eV in the Ar$^{2+}$ yield corresponds to the SOWP oscillation, and the peak at approximately 1.3 eV in the N$^{+}$/N$^{+}$ yield corresponds to the expected $X-A$ beating in the N$_{2}^{+}$ cation. The broad peak in both species spanning 1.3-1.7 eV corresponds to OI caused by the overlapping pulses. However, even though the $I_p$ for both ions are very similar, it is clear that the OI effect manifests very differently in the two species. The amplitude of the FFT spectrum of the N$^{+}$/N$^{+}$ and Ar$^{2+}$ signals are distinct, corroborating the experimental results.
In particular, around the broad OI peak in Fig.~\ref{NpNp_Arpp_exp}b, the amplitudes of the two species are clearly different.
This indicates why subtraction schemes using Ar$^{2+}$ to remove OI background from the N$^{+}$/N$^{+}$ signal are not possible.

Since the simulated signals of N$^{+}$/N$^{+}$ ions in coincidence and Ar$^{2+}$ ions are so different even though their $I_p$ are similar, a question arises of what features of the two species are causing the measured differences. To eliminate complicated effects arising from the electronic structure of N$_2^+$, we compare the OI signals between Kr and Ar.
Figure~\ref{oi_sim_compare}b displays the FFT spectrum of the Ar$^{2+}$ and Kr$^{2+}$ signals.
The peaks at 0.18 and 0.67 eV, which correspond to spin-orbit energy separation of Ar$^{+}$ and Kr$^{+}$, are clearly observed.
However, the amplitude and phase of the OI signal at around 1.7 eV is significantly different between Ar$^{2+}$ and Kr$^{2+}$.
This suggests that even for atomic ions with a similar (albeit energetically distinct) set of electronic dynamics, their OI effects are distinct.

To reveal the dominant factors that determine how OI manifests in noble gas atoms, we consider an artificial Kr atom, in which the spin-orbit energy separation of Kr$^+$ is replaced with that of Ar$^+$, while the first and second $I_p$ remain fixed.
We compare the FFT spectrum of the artificial Kr$^{2+}$ with Ar$^{2+}$ in Fig.~\ref{oi_sim_compare}b.
The amplitude of artificial Kr$^{2+}$ at 0.18 eV is larger than that of Ar$^{2+}$ since Kr's $I_p$ are lower.
However, the amplitude and phase of their OI signals from 1.3 to 1.7 eV are very similar, indicating that the ionic structure, specifically the spin-orbit splitting, plays a dominant role in the mechanism of OI.

To further investigate the role of ionic structures in OI, we simulate the pump-probe schemes for fixed nuclei N$_2$ when dipole couplings between N$_2^+$ or N$_2^{2+}$ are switched off.
Figure~\ref{oi_sim_compare}c compares the simulated signals. 
When the laser coupling for N$_{2}^{+}$ is turned off, while the phases remain about the same, the amplitudes of the FFT spectrum change significantly. In particular, the amplitude of the 1.6 eV OI peak in the case of $\vec{d}^{(1)} = 0$ is about a factor of 3 larger than the case with all laser couplings switched on.
In contrast, removing the dication couplings instead has a much smaller effect on the FFT spectrum, indicating that the dipole coupling in the cation is much more important than the dication. The amplitudes and phases from 1.3 to 2.0 eV are similar to the case with all laser couplings switched on.
This simulation demonstrates that even with the same ionic levels, changing the ionic dipole moments will change the OI effect significantly.

These calculations suggest that OI effects manifest very differently in each system, with a strong dependence on the electronic structure of the species. Unfortunately, this means that the subtraction schemes that were alluded to earlier for attempting to remove OI contributions are not possible, because each species always has a different electronic structure. The presence of a coherence signal near the laser frequency will itself modulate the OI in a complicated way that makes it impossible to replicate with another system. Beatings at significantly different frequencies than the laser frequency may be possible to resolve in the presence of an OI peak, but they are not sensitive to laser couplings, a necessary component for strong-field control of electronic coherence.
Therefore, to pave the way for strong-field control of chemical reactions, it is necessary to perform pulse cleaning on the few-cycle intense IR pulses to remove the satellite structures and pedestal.

\onecolumngrid

\newpage
\twocolumngrid

\section{Role of Nonsequential Double Ionization~\label{sec6}}
After ruling out the possibility of background subtraction of the OI signal in one species using the OI signal in another species, a question remains: can we perform background subtraction in the dication signal of one species using its single-cation signal?
To answer this question, we investigate the simulated and measured Ar$^+$ and Ar$^{2+}$ yields as a function of interpulse delay.

\begin{figure}[htb]
	\centering
	\includegraphics[width=\linewidth]{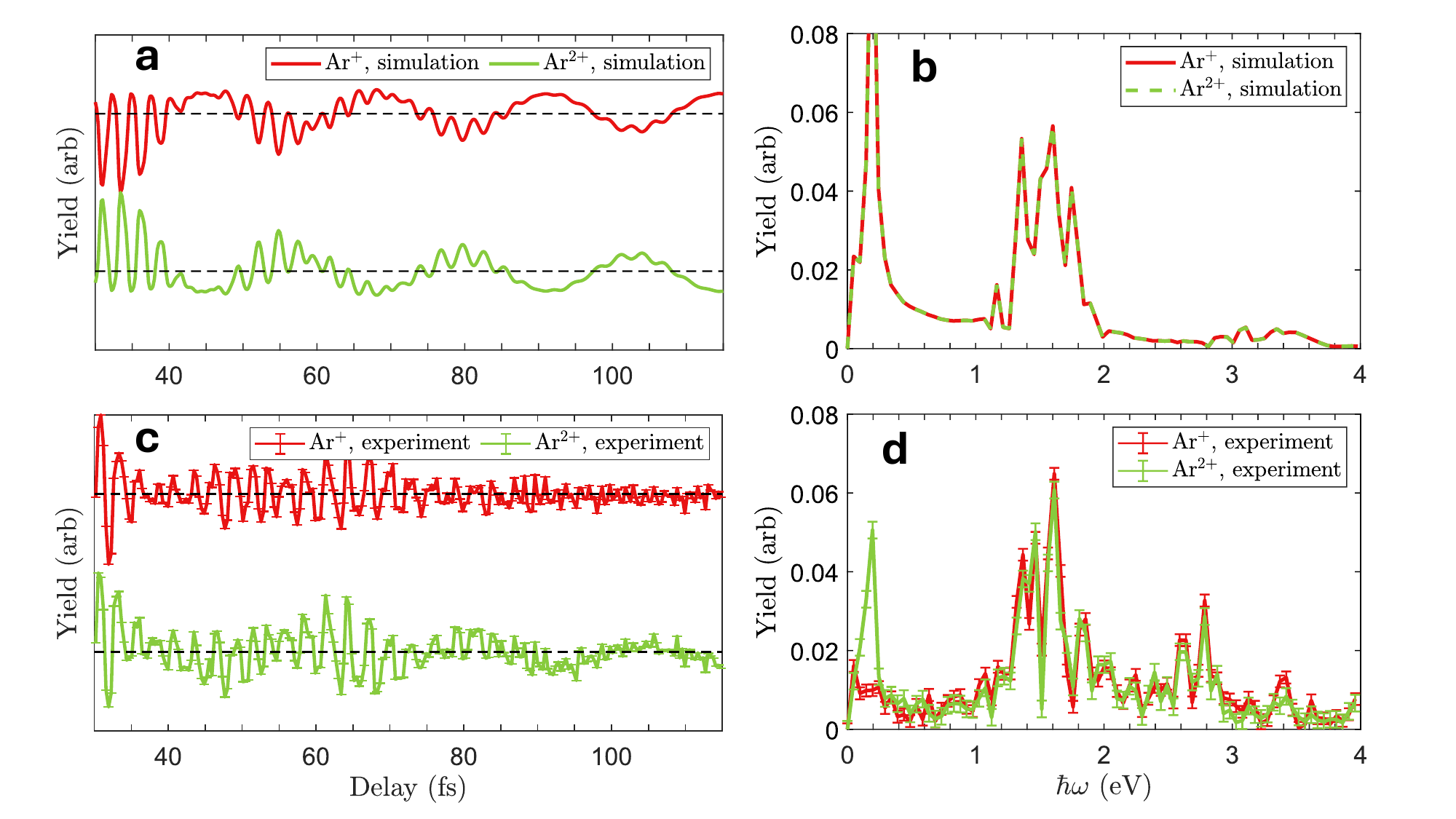}
	\caption{ Comparison of Ar$^{+}$ and Ar$^{2+}$ yield in both simulation and experiment. \textbf{(a)} Simulated yield of Ar$^{+}$ and Ar$^{2+}$ ions versus delay, demonstrating the expected $\pi$ phase shift between the two species \textbf{(b)} FFT amplitude of the data in (a) demonstrating the expected identical frequency signal of both species \textbf{(c)} experimental yield of Ar$^{+}$ and Ar$^{2+}$ ions versus delay \textbf{(d)} FFT amplitude of the data in (c). Error bars on the experimental data reflect statistical uncertainties.}
	\label{self_subtraction}
\end{figure}

Figure~\ref{self_subtraction}a displays Ar$^{+}$ and Ar$^{2+}$ yield simulated using the DM-SDI model with the retrieved laser pulse parameters used in Fig.~\ref{sim_exp_compare}.
According to Eqs.~\eqref{eq:argon1} and \eqref{eq:argon2}, enhancements in Ar$^{2+}$ ionization at specific delays caused by the SOWP oscillation are matched by an equivalent decrease in the Ar$^{+}$ signal simply due to depletion. 
This means that in the SDI regime, the Ar$^{+}$ and Ar$^{2+}$ yield always has a $\pi$ phase shift.
This effect is seen clearly in the time domain data in Fig.~\ref{self_subtraction}a, where depletions in the Ar$^{+}$ signal occur at the same delay as enhancements in the Ar$^{2+}$ signal. This effect is also shown in Fig.~\ref{self_subtraction}b, which shows the FFT amplitude of the data in panel (a). The measured amplitudes are identical across all frequencies for both species since they differ only by a $\pi$ phase shift.

Figure~\ref{self_subtraction}c shows the measured Ar$^+$ and Ar$^{2+}$ signal at different delays and their FFT spectra.
The peaks and troughs of Ar$^+$ and Ar$^{2+}$ signals are well aligned, suggesting their OI signals are in phase, instead of having a $\pi$ phase shift as in the simulation.
In addition, the slow varying envelope corresponding to the SOWP oscillation is not visible in the Ar$^+$ signal.
These observations are further confirmed in the FFT spectrum in Fig.~\ref{self_subtraction}d.
The 0.18 eV peak in the Ar$^+$ signal is significantly weaker than in the Ar$^{2+}$ signal, while the OI peaks have about the same position and magnitude.

The disagreement between simulation and experiment in Fig.~\ref{self_subtraction} can be explained by the contribution to the Ar$^{2+}$ signal from nonsequential double ionization (NSDI) by the probe pulse at different focal volumes.
NSDI is driven by the tunnel ionization of Ar and the $(e,2e)$ recollision ionization or the recollision excitation with subsequent ionization of Ar$^+$~\cite{Bergues2012_Attosecond}.
While the first step is highly sensitive to OI, the second step is insensitive to changes in the scattering energy and, therefore, less sensitive to OI.
In addition, the depletion of the ionic population is negligible in NSDI.
The combined effect is that the NSDI yield at different delays changes in phase with the Ar$^+$ yield.
Since the recollision process is insensitive to the spin-orbit structure~\cite{Liu2024_Exploring}, the impact of SOWP oscillation is minimal, and the contribution from NSDI to the FFT spectrum of Ar$^{+}$ and Ar$^{2+}$ at 0.18 eV is negligible.

The overall contribution from NSDI can be inferred from the modulation depth in the measured Ar$^{2+}$ signal shown in Fig.~\ref{sim_exp_compare}. The oscillation amplitude of the measured Ar$^{2+}$ signal is about 20 times weaker than that in the simulation, which neglects the contribution from NSDI. Therefore, it is likely that NSDI plays a dominant role in contributing to the Ar$^{2+}$ signal. While the yield of NSDI is known to be less than that of SDI, a possible reason for its overall dominance is that weaker peak intensities carry greater weights in focal volume averaging. This dominance explains why the measured OI signals in Ar$^{2+}$ and Ar$^+$ are in phase, and why the FFT spectrum of the measured Ar$^+$ yield has a significantly weaker signal at 0.18 eV compared to the Ar$^{2+}$ signal.

In summary, subtracting the OI signals of the dication formed in the focal volume in the SDI regime using its ion signal is clearly unfeasible because the OI signals of ion and dication primarily originate from the focal volume in the NSDI regime. Our investigation shows that NSDI is likely to significantly diminish the contrast of the electronic coherence signal. One solution is to tightly focus the pump pulse while loosely focusing the probe pulse into the vacuum chamber. This ensures that the focal volume in the NSDI regime is outside the focal volume of the pump pulse, allowing the overall NSDI yield to serve as the background signal, which can be subtracted using the probe-only measurement.
We emphasize that pulse cleaning is still necessary despite the focal volume control.
Otherwise, the pump pulse could still interfere with the probe pulse in the NSDI focal volume, producing OI signals that cannot be subtracted using the probe-only measurement.
While the use of elliptical polarization could also suppress NSDI, it would simultaneously diminish the electronic coherence and may therefore not be suitable for the present study.

\onecolumngrid

\begin{figure}[htb]
	\centering
	\includegraphics[width=\linewidth]{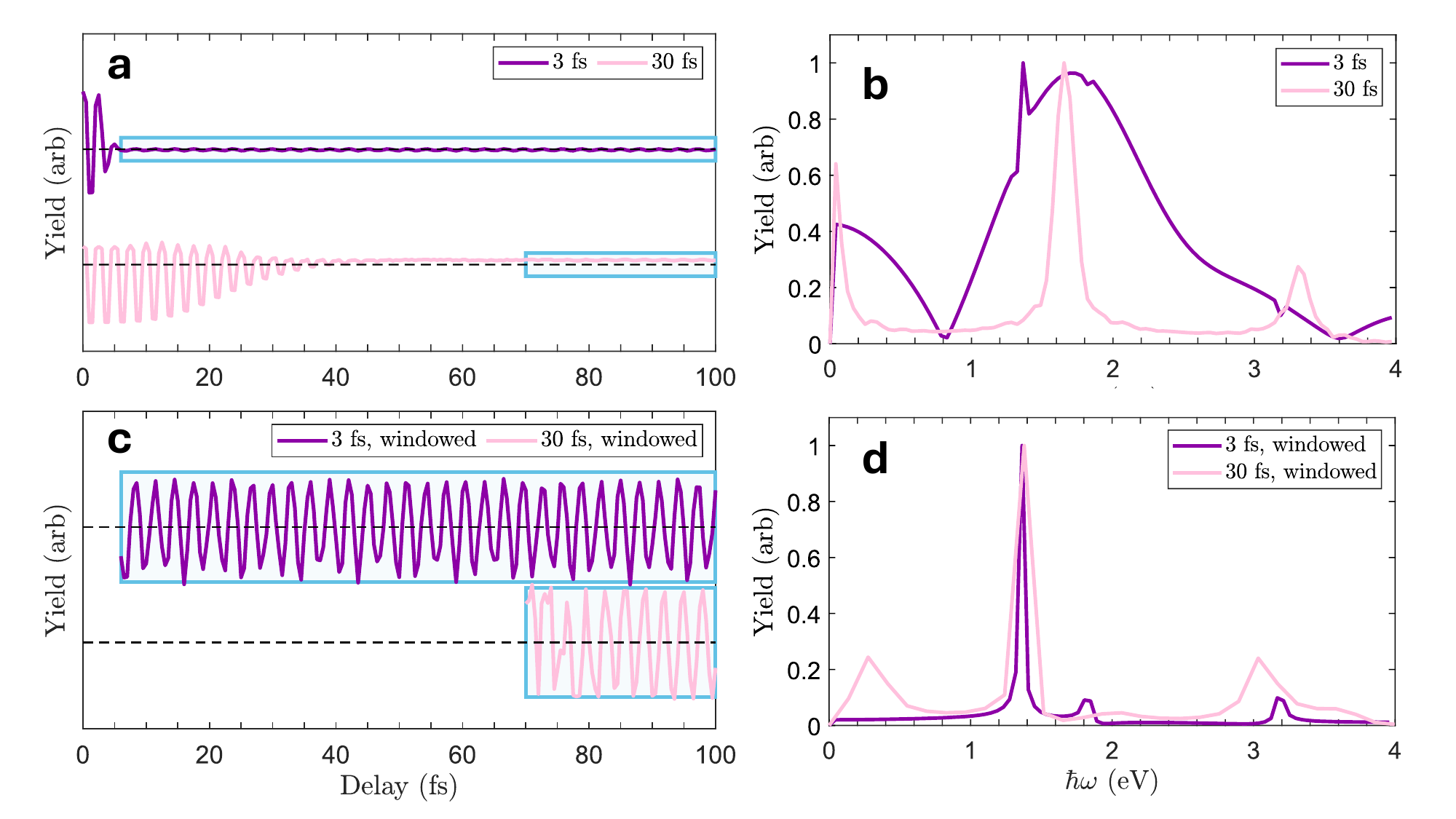}
	\caption{
	Simulated N$^{+}$/N$^{+}$ yield from fixed-nuclei N$_2$ using Gaussian pulses of 3 fs and 30 fs (FWHM in intensity) with a fixed carrier envelope phase of zero. \textbf{(a)} N$^{+}$/N$^{+}$ yield scanned over an interpulse delay of 0-100 fs. \textbf{(b)} FFT amplitude of all data in (a). \textbf{(c)} time windowed data in the blue boxes in panel (a), where optical interference effects are negligible. \textbf{(d)} FFT amplitude of the data in (c). }
	\label{pulse_duration}
\end{figure}

\newpage
\twocolumngrid

\section{Optical Interference Effects in an Ideal Scenario~\label{sec7}}

In the ideal scenario, the envelope of the single- or few-cycle pump and probe pulses are Gaussian, their carrier envelope phases are stabilized and set to zero, and the contribution to the dication yield from nonsequential double ionization is subtracted by the focal volume control.
Will optical interference (OI) effects still take place in strong field dissociative sequential double ionization (DSDI) spectroscopy for electronic coherence?
How do the OI effects in the ideal scenario depend on the laser bandwidth?
To answer these questions, we simulate the strong field pump-probe scheme for fixed-nuclei N$_2$ with transform-limited Gaussian pulses with a duration of 3 fs and 30 fs (FWHM in intensity).
Other laser parameters are described in Sec.~\ref{sec3.2}.
The pulse duration of 3 fs is used, as it could be realized with novel pulse compression techniques~\cite{Tsai2022_Nonlinear}.
The nuclei are frozen throughout because the pump and probe pulses must overlap to simulate the OI effect, while the current DM-SDI model cannot treat the laser-molecule interaction with moving nuclei.
The interpulse delay is scanned from 0 to 100 fs, so the FFT spectra include data where the two pulses are completely overlapped in time, generating maximum OI.

In Fig.~\ref{pulse_duration}a, simulated N$^{+}$/N$^{+}$ yield at different delays using the 3 fs and 30 fs Gaussian pulses are compared.
When the interpulse delay is less than twice of the pulse duration, results from both pulses have a very strong OI contribution.
Figure~\ref{pulse_duration}b shows the FFT spectra calculated using the entire delay range.
The narrow OI peak at 1.6 eV is clearly visible in the case of 30 fs pulses.
The two peaks at around 0.1 and 3.2 eV are due to the beating between the laser frequency and the $X-A$ coherence (1.3 eV). These peaks are shifted by about -0.2 and +0.2 eV because of the ac Stark shift.
In the case of 3 fs pulses, the OI peak and its satellite peaks are centered at the same frequencies as in the case of 30 fs pulses, but they are much broader.
Despite the OI, the $X-A$, $A-B$, and $X-B$ beating signals at 1.3, 1.8, and 3.2 eV are still visible.

In Fig.~\ref{pulse_duration}c, a time window is applied to each simulated yield to remove contributions from pulse delays for which the two pulses have significant temporal overlap. In the 3 fs case, this means only including the simulation from interpulse delays of 6-100 fs in the FFT signal. When this time window is applied for the 3 fs pulse, the OI effects are completely removed from the FFT spectrum in Fig.~\ref{pulse_duration}d, and the 1.3, 1.8, and 3.2 eV beatings are sharp. In the 30 fs case, this means only including the simulation from interpulse delays of 70-100 fs in the FFT signal. When this is done, the OI contribution is almost eliminated, and the primary peak shifts from the laser frequency to the 1.3 eV $X-A$ coherence signal. However, the beating between the laser frequency and the $X-A$ coherence is still visible due to the residual OI effects from delays of 70 to 80 fs.

Our results show that, in the ideal scenario, OI effects can be mitigated simply by filtering the data in the time domain, where the pump and probe pulses are delayed by at least more than twice the pulse duration.
To resolve the onset of electronic decoherence due to nuclear motion, the the ideal minimum delay is around 3 fs~\cite{Arnold2017_Electronic}. Our findings suggest that the ideal pulse duration for DSDI spectroscopy is about 1.5 fs. Therefore, cleaning an IR pulse generated by soliton self-compression~\cite{Brahms2020_Infrared} could significantly enhance the use of DSDI spectroscopy to probe electronic decoherence in molecules.
Finally, we find that the effectiveness of mitigating OI effects mainly depends on the narrowness of the pulse in the time domain rather than in the frequency domain. Although the broad OI peak of the 3 fs pulse may mask the electronic coherence signal, its short pulse duration allows the OI effect to be effectively filtered. In contrast, while the OI peak of the 30 fs pulse is sharp, filtering the OI effect proves to be more challenging.

\section{Summary and Outlook~\label{sec8}}
The results of this joint theory-experiment study highlight the significant role that optical interference (OI) plays in strong field dissociative sequential double ionization (DSDI) spectroscopy for electronic coherence, with unique effects in different atomic and molecular systems. 
The origin of the OI effects is the overlap between the pump and probe pulses.
For ideal Gaussian pulses, we theoretically demonstrated that the OI effects can be mitigated by filtering the signal in the time domain so that the pump-probe overlap does not affect the electronic coherence signal.
However, experimentally, the few-cycle infrared pulses generated by nonlinear compression often contain satellite pulses and pedestals.
The interference between these structures and the main pulse leads to signals lasting longer than 100 fs that wash out the target's electronic coherence signal.
Time filtering is no longer feasible because molecular rotation may lead to dephasing at longer delays.
While background subtraction of the OI signal with species that share similar ionization potentials is an appealing option, using N$_2$ and Ar as examples, we theoretically and experimentally demonstrated that OI cannot be eliminated with this approach, as the OI effect strongly depends on the ionic structure of the species involved.

By comparing simulated results with the experiment, we further determined that the background subtraction of the OI signal in a dication using its ion signal is not feasible because their OI signals originate from different focal volumes.
We also revealed that nonsequential double ionization (NSDI) by the probe pulse at the outer focal volume considerably contributes to the dication yield, which carries the OI signal but not the electronic coherence signal.
As a result, NSDI significantly reduces the overall amplitude of coherence signals but increases the amplitude of the OI signals.

Our study concludes that to perform strong field DSDI spectroscopy for electronic coherence, the necessary steps in future experiments are: 

(1) Pulse cleaning to eliminate satellite pulses and suppress the pedestal, thereby mitigating the OI effects~\cite{Jullien2006_Highly, Smijesh2019_Contrast}.

(2) Stabilization of the carrier envelope phase to suppress background noise in the FFT spectra.~\cite{Cerullo2011_Fewopticalcycle, Moon2006_Reductionc, Adachi2004_Quasimonocyclic}

(3) Focal geometry control to set the outer focal volume of the probe pulse in the NSDI regime outside the focal volume of the pump pulse where strong field ionization takes place, allowing NSDI signals to be eliminated as background signals.


The experimental guidelines presented in this work can bridge the gap between strong field theory and experiment and create a plethora of research opportunities in charge migration and transfer using table-top strong field DSDI spectroscopy.
The use of the DM-SDI model~\cite{Yuen2022_Densitymatrix, Yuen2023_Coherence, Yuen2024_Probing, Yuen2024_Theory} allows us to simulate the dication yield in a pump-probe scheme with realistic laser pulses so that the OI signals in the observables can be qualitatively understood.
We note such simulations are particularly challenging for state-of-the-art first-principles approaches.
This is because one must either account for the double continuum electrons under a highly intense laser field or effectively absorb the enormous fluxes of ionized electrons in a simulation box while resolving the final dication electronic states to map to the experimental observables.
As noted in Ref.~\cite{Yuen2024_Theory}, the observables from the ideal strong field DSDI spectroscopy can be interpreted without simulating the DSDI probe process.
Moving forward with pulse cleaning techniques for compressed IR pulses and with simulations using the DM-SDI model,  future simulations could be greatly simplified, allowing the pump and probe processes and the interpulse dynamics to be treated separately, as discussed in Ref.~\cite{Yuen2024_Probing, Yuen2024_Rotation}.
We anticipate that in the near future, first-principles approaches may become more integrated with strong field experiments by providing accurate descriptions of the population and coherence in pumped states and interpulse molecular dynamics. This integration will facilitate the investigation of charge migration and transfer in various molecules.

As pulse cleaning and strong field probing techniques advance, there is the potential for a major breakthrough in strong field ultrafast chemistry control. By adding a control pulse between the pump and probe pulses where OI effects are suppressed, we can explore the complex mechanisms of strong field control of electronic coherence, find the optimal target molecules for manipulation, and reveal the important links between electronic coherence and the reactivity and functionality of molecules. A new and exciting area in quantum technology is developing, and this paper lays out a potential roadmap to achieve it.

\begin{acknowledgments}
E.W., A.J.H., C.C., I.G., A.M.G., and P.H.B. were supported by the National Science Foundation. A.J.H. was additionally supported under a Stanford Graduate Fellowship as the 2019 Albion Walter Hewlett Fellow. A.M.G. was additionally supported by an NSF Graduate Research Fellowship. C.D.L. and C.H.Y. were supported by Chemical Sciences, Geosciences and Biosciences Division, Office of Basic Energy Sciences, Office of Science, U.S. Department of Energy under Grant No. DE-FG02-86ER13491.

The data that support the findings of this work are openly available \cite{Weckwerth2025_Data}.
\end{acknowledgments}

\appendix
\section*{Appendix}
\renewcommand{\thefigure}{A\arabic{figure}}
\setcounter{figure}{0}

\section{Pulse Characterization and Retrieval}
\label{app:pulse}

The pulses are characterized using the dispersion scan method as discussed in the Experiment section (Section~\ref{sec3.1}), which works by overcompensating for the positive chirp accrued in the spectral broadening from the hollow core fiber using extra bounces on negatively chirped mirrors and then scanning over a range of glass insertion to introduce additional positive chirp until the pulse is optimally compressed~\cite{Miranda2012_Simultaneous,Sytcevich2021_Characterizing}. The minimum pulse duration is found by measuring the SHG yield as a function of glass insertion and comparing to the expected yield from an estimated spectral intensity $I(\lambda)$ and phase $\phi(\lambda)$, then adjusting the values of $I(\lambda)$ and $\phi(\lambda)$ until they converge to a value that agrees with the measurement.

\begin{figure}[tb]
	\centering
	\includegraphics[width=\linewidth]{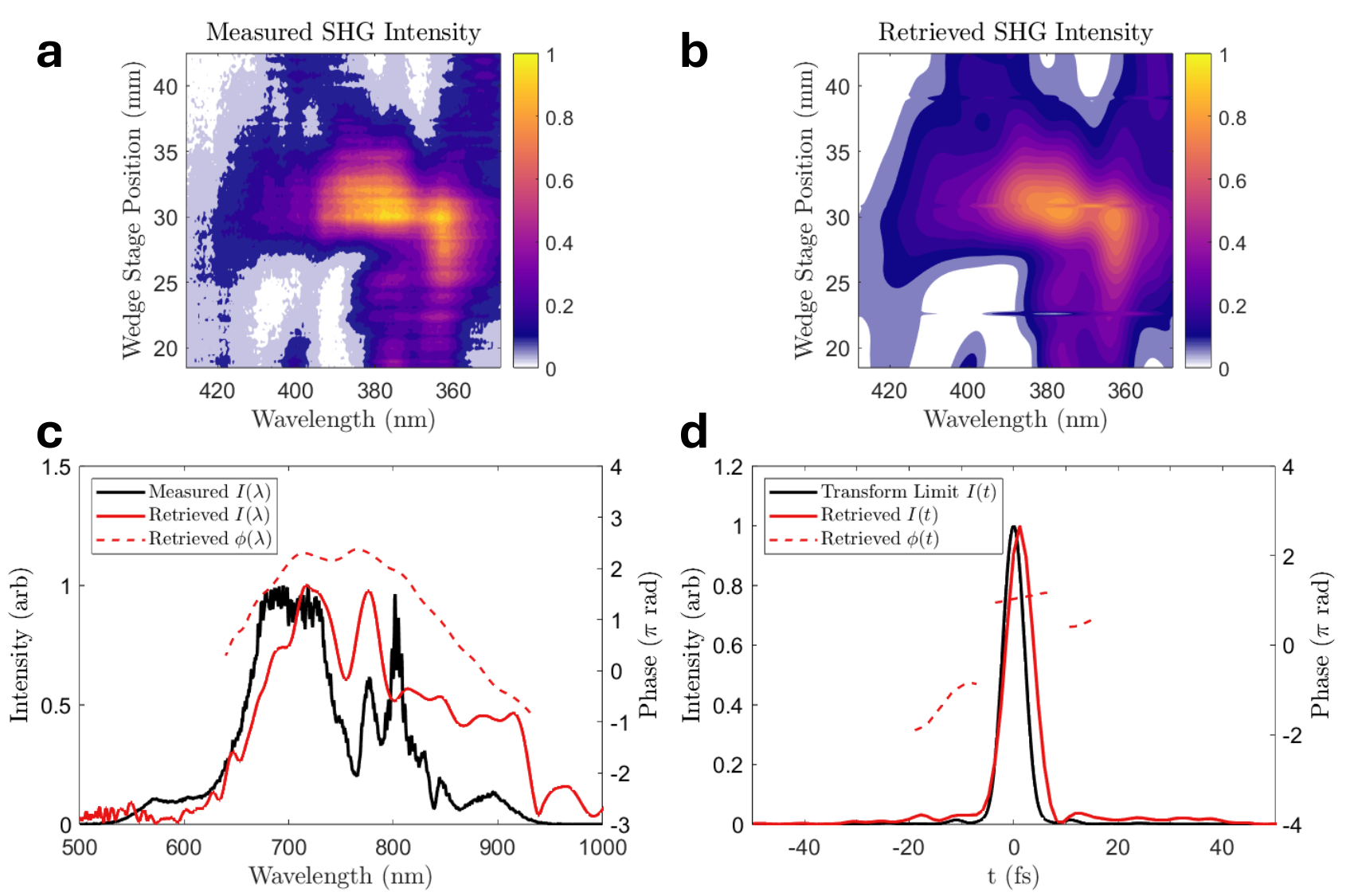}
	\caption{Dispersion scan measurement for experimental pulses and retrieved pulse used in simulations. \textbf{(a)} Measured and \textbf{(b)} retrieved SHG yield as a function of both wavelength and glass insertion. \textbf{(c)} Measured fundamental spectrum (black) and retrieved spectrum and phase from the dispersion scan (red). \textbf{(d)} The transform-limited temporal intensity envelope based on the measured fundamental spectrum (black) and retrieved temporal intensity envelope of the pulse from the dispersion scan (red).}
	\label{dscan}
\end{figure}

The result of the dispersion scan is a retrieved $I(\lambda)$ and $\phi(\lambda)$ which can then be used to extract information about the temporal envelope of the electric field of the pulse. Figure~\ref{dscan}a (and b) show the measured (and retrieved) SHG intensity as a function of wavelength and glass insertion. Figure~\ref{dscan}c shows the measured fundamental spectrum (black) and the retrieved spectrum used to reproduce the SHG yield, both intensity (red solid) and phase (red dashed). The measured spectrum decreases with increasing wavelength due to the nonuniform response of the spectrometer, but this does not affect the retrieval. Figure~\ref{dscan}d shows the Fourier-transform- limited pulse based on the measured fundamental spectrum (black, FWHM 4.6 fs) and the retrieved $I(t)$ from the dispersion scan, which is a more realistic approximation of the pulse duration, including higher-order phase on the pulse (red, FWHM 6.5 fs). The effects of the imperfect higher-order phase compensation are clear from the retrieved pulse duration and asymmetric shape in the time domain, which becomes important for OI effects.

\onecolumngrid

\newpage
\twocolumngrid

\begin{figure}[tb]
	\centering
	\includegraphics[width=\linewidth]{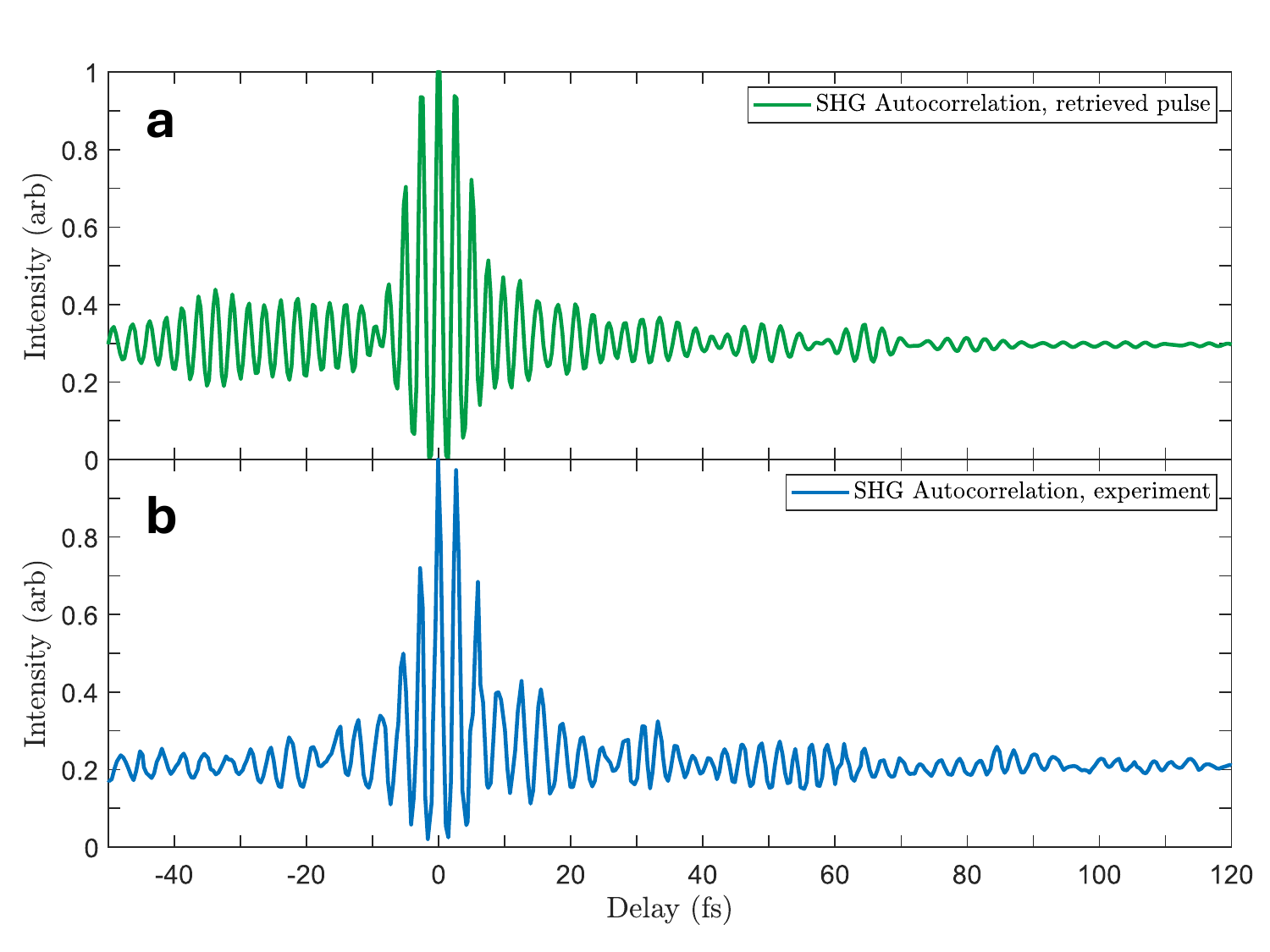}
	\caption{Comparison of retrieved pulse and measured pulse using interferometric autocorrelation \textbf{(a)} Simulated SHG intensity autocorrelation using retrieved $E(t)$ from dispersion scan. \textbf{(b)} Experimentally measured SHG autocorrelation, which is the same signal compared in Figure~\ref{fft_NpNp_shg}.}
	\label{autocorr}
\end{figure}

The retrieved $E(t)$ from the dispersion scan is used in the simulations to recreate the experimental conditions as accurately as possible. To ensure that an accurate retrieval is used, we can compare it to another method of pulse characterization discussed in the Experiment section, the SHG intensity autocorrelation, which is an interferometric technique that scans the interpulse delay and detects the interference of the pulses in an SHG crystal. An SHG autocorrelation can be modeled with the retrieved pulse by scanning two copies of $E(t)$ against each other, with one a factor of 3.1 smaller to match the experiment. The resulting retrieved autocorrelation is shown in Figure~\ref{autocorr}a, and compared with the experimentally measured autocorrelation in~\ref{autocorr}b. The two signals are qualitatively similar over the positive delays scanned in the experiment, indicating that the retrieval is a reasonable approximation of the experimental pulses.

\section{Systematic stage motion frequency}
\label{app:stage}
\begin{figure}[htb]
	\centering
	\includegraphics[width=\linewidth]{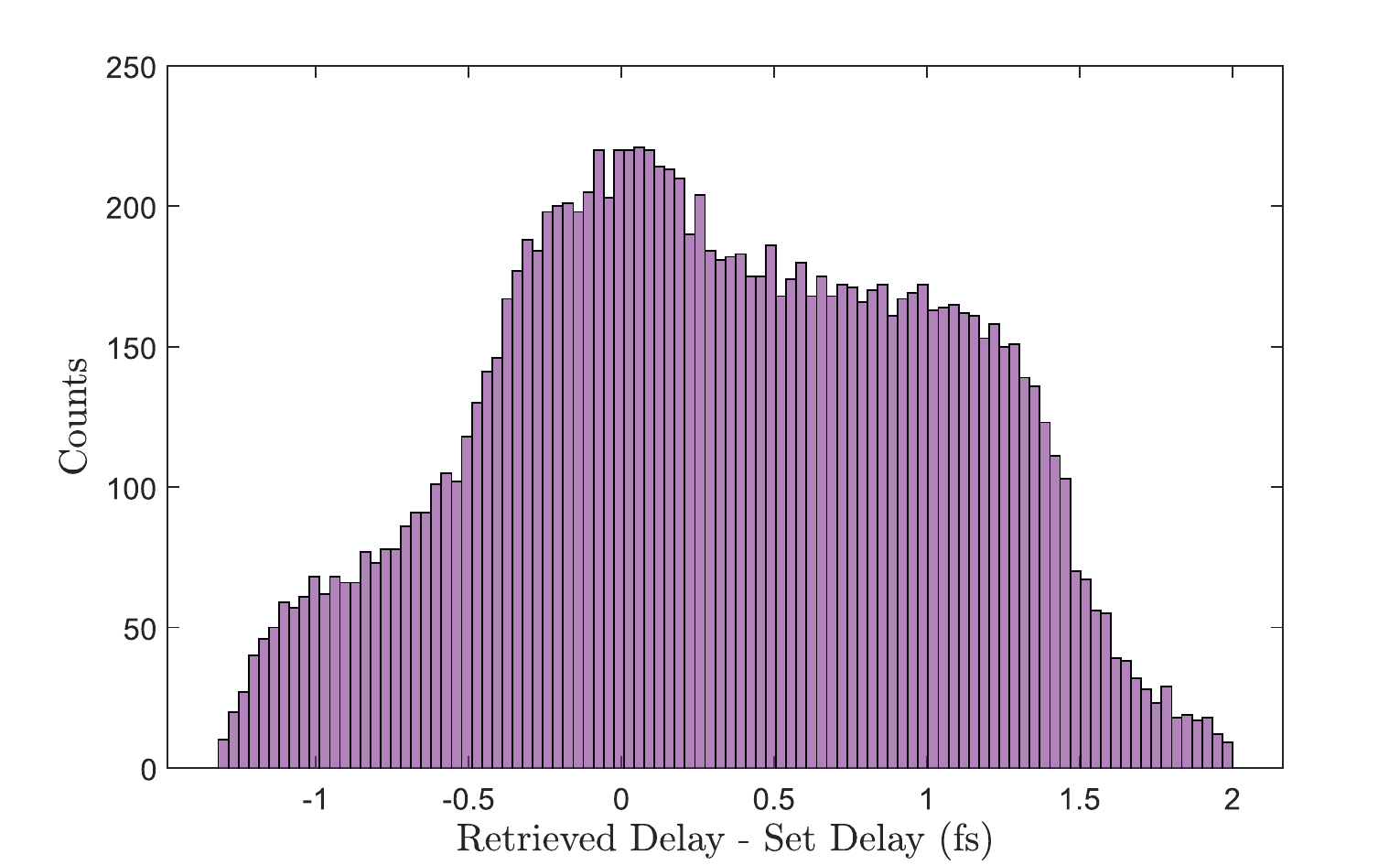}
	\caption{Comparison of retrieved delay from spectral interference to set delay for all stage positions, demonstrating the error in the arrival position of the stage }
	\label{stagehist_together}
\end{figure}

The generation of interpulse delays using a translation stage produces some experimental challenges which manifest in the data. The stage creating the interpulse delays was nominally set to move in 0.5 fs steps across the range of measured delays. Due to mechanical errors like hysteresis and backlash, the arrival position of the stage is not always equal to the assigned position, so the delays can vary from the set point by as much as a few femtoseconds. The actual delays used in the experiment are calculated using the spectral interference between the two pulses as described in the Experiment section, recovering the actual delays with much more precision than the stage can report, so the arrival position of the stage does not need to match the intended position exactly. The data in this experiment was collected over many iterations of the stage motion, so over time the gaps between the set delay positions are filled in to sample the entire range of delays. This is shown in Figure~\ref{stagehist_together}, which shows a histogram of the difference between the retrieved delay from spectral interference and the set delay assigned to the stage for all stage positions throughout the experiment. The position of the stage can vary from its set point by more than 1 fs, but this is corrected in the analysis by using the much more accurate retrieved delays.

\begin{figure}[htb]
	\centering
	\includegraphics[width=\linewidth]{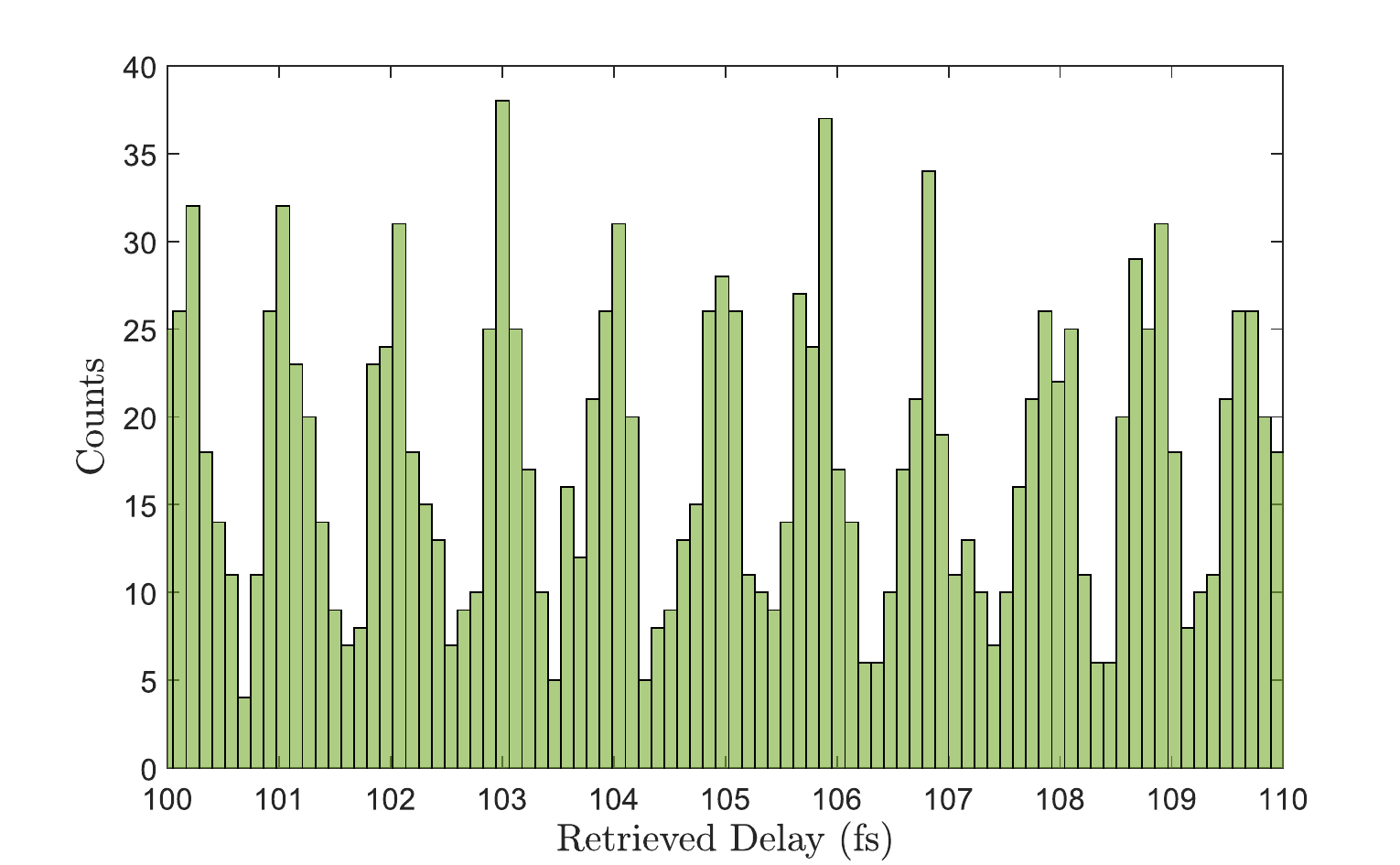}
	\caption{Histogram of retrieved delays from spectral interference for a range of stage positions, demonstrating the modulation in delay sampling}
	\label{stagehist_spread}
\end{figure}

However, this effect still manifests in the data in a small way. Comparison of many different ion channels measured in the experiment shows a peak at approximately 4.25 eV which appears similarly in all of the channels. This frequency corresponds to the Nyquist frequency for time domain data in 0.5 fs steps, indicating a connection to the sampling of delays in the data. In Figure~\ref{stagehist_spread}, the distribution of retrieved delays for a small sample of the delay range demonstrates a clear modulation.

This problem is mostly alleviated by normalizing every channel histogrammed in delay by the number of total laser shots in that delay bin. However, for channels with lower counting statistics, the variation in counts for some of the less sampled delays could be large enough that the modulation is still seen. This results in a systematic frequency peak that appears to some degree in all channels but especially in channels with fewer statistics.

The existence of this additional frequency peak does not limit the conclusions that can be drawn from the data, since it appears at an energy that does not interfere with any other expected signals. However, for clarity in analysis and comparison of normalized FFT spectra, we have eliminated the single peak at 4.25 eV from all experimental channels. It is also removed from the time domain data by using an inverse fast Fourier Transform to reproduce the delay-resolved data without the 0.5 fs modulation.

\section{KER-resolved analysis}
\label{app:KER}
The N$^{+}$/N$^{+}$ channel can be further divided based on the final dicationic state produced using the measured Kinetic Energy Release (KER). The expected signal due to electronic coherences changes with KER because each final dication state is coupled with a particular set of cationic coherences~\cite{Yuen2024_Probing}. Since the relative intensity of the strongest electronic coherence signal is expected to be different for each dication state, filtering by dication state could improve the signal contrast in one of the filtered channels.

\begin{figure}[tb]
	\centering
	\includegraphics[width=\linewidth]{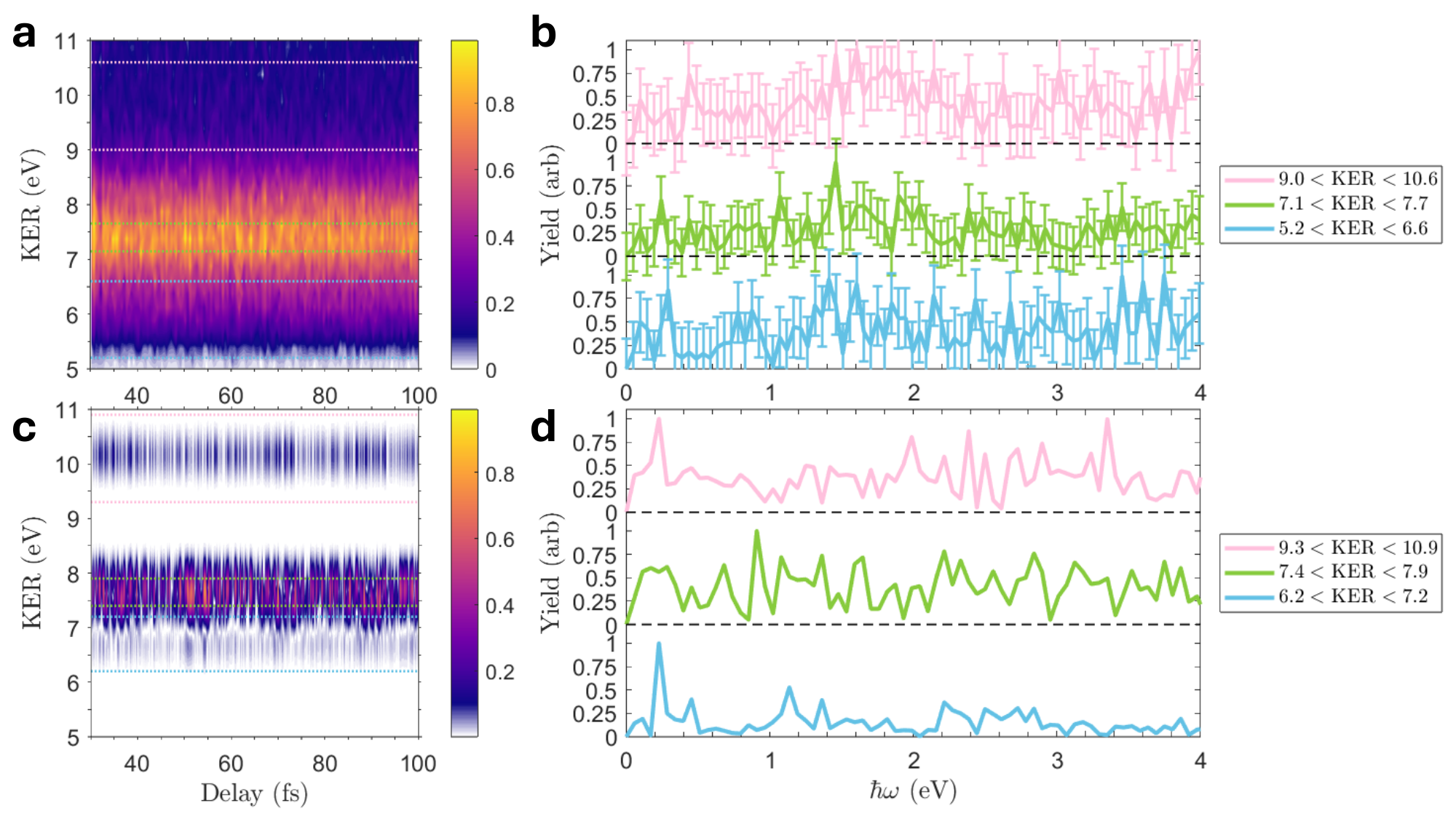}
	\caption{KER-resolved yield vs delay and FFT amplitude for N$^{+}$/N$^{+}$ ion coincidence channel. \textbf{(a)} Experimental and \textbf{(c)} simulated yield as a function of KER and delay, with horizontal dotted lines demonstrating KER ranges selected from a small distribution around peaks in the KER yield corresponding to different dication channels. \textbf{(b)} Experimental and \textbf{(d)} simulated FFT amplitude for N$^{+}$/N$^{+}$ ions within the ranges plotted in (a) and (b). Error bars on the experimental data reflect statistical uncertainties.}
	\label{ker_2d}
\end{figure}

Experimental N$^{+}$/N$^{+}$ coincidence ion yield as a function of KER and delay is shown in Figure~\ref{ker_2d}a. The dissociative states of N$_{2}^{2+}$ correspond to KERs of 6.5, 7.5, and 10.0 eV, which manifest as horizontal stripes in the 2D plot, have been identified in Refs.~\cite{Yuen2022_Densitymatrix, Jia2024_Improved}. Peaks in the KER distribution corresponding to one of these three energies were selected with a Gaussian fit, and a small range symmetric about the peaks was sampled such that each range has a similar amount of data. Horizontal dotted lines show the limits of these selected KER regions.

An FFT spectrum of the data in the marked regions from Figure~\ref{ker_2d}a is shown in Figure~\ref{ker_2d}b, with each lineout corresponding to one of the three listed dication states. The three signals, separated by KER, all largely share similar structure in both the time- and frequency-domain, as seen in Figs~\ref{ker_2d}a and~\ref{ker_2d}b. None of the FFT amplitudes in the three different regions show obvious coherence beatings near the expected frequencies.

Additionally, the FFT signals in Figure~\ref{ker_2d}b show that the broad OI peaks manifest differently in the three different channels, indicating that either OI has a different effect in the three different KER channels or there are some non-optical beatings contributing to the signal. The observed differences in OI manifestation demonstrated in Figure~\ref{oi_sim_compare} suggest that different pathways to ionization respond to the presence of a laser field in various ways. Therefore, the different formation pathways of various N$_{2}^{2+}$ dications could explain the variation in the resulting FFT signal for filtered KER-specific channels seen in Figure~\ref{ker_2d}b.

To differentiate between these conclusions, we extract the KER spectrum from Fig.~\ref{oi_sim_compare} for fixed-nuclei N$_2$, as shown in Figure~\ref{ker_2d}c and~\ref{ker_2d}d. The three distinct KER channels are much more pronounced and well separated in the simulated data. KER lineouts are selected in the same way as the experimental data, and the frequency spectrum for the chosen KER regions is shown in Figure~\ref{ker_2d}d. The resulting signals are similarly noisy and do not have consistent spectral character in the OI region. This confirms that OI effects vary with different ionization pathways.

\onecolumngrid

\newpage
\twocolumngrid


\begin{thebibliography}{60}%
\makeatletter
\providecommand \@ifxundefined [1]{%
 \@ifx{#1\undefined}
}%
\providecommand \@ifnum [1]{%
 \ifnum #1\expandafter \@firstoftwo
 \else \expandafter \@secondoftwo
 \fi
}%
\providecommand \@ifx [1]{%
 \ifx #1\expandafter \@firstoftwo
 \else \expandafter \@secondoftwo
 \fi
}%
\providecommand \natexlab [1]{#1}%
\providecommand \enquote  [1]{``#1''}%
\providecommand \bibnamefont  [1]{#1}%
\providecommand \bibfnamefont [1]{#1}%
\providecommand \citenamefont [1]{#1}%
\providecommand \href@noop [0]{\@secondoftwo}%
\providecommand \href [0]{\begingroup \@sanitize@url \@href}%
\providecommand \@href[1]{\@@startlink{#1}\@@href}%
\providecommand \@@href[1]{\endgroup#1\@@endlink}%
\providecommand \@sanitize@url [0]{\catcode `\\12\catcode `\$12\catcode `\&12\catcode `\#12\catcode `\^12\catcode `\_12\catcode `\%12\relax}%
\providecommand \@@startlink[1]{}%
\providecommand \@@endlink[0]{}%
\providecommand \url  [0]{\begingroup\@sanitize@url \@url }%
\providecommand \@url [1]{\endgroup\@href {#1}{\urlprefix }}%
\providecommand \urlprefix  [0]{URL }%
\providecommand \Eprint [0]{\href }%
\providecommand \doibase [0]{https://doi.org/}%
\providecommand \selectlanguage [0]{\@gobble}%
\providecommand \bibinfo  [0]{\@secondoftwo}%
\providecommand \bibfield  [0]{\@secondoftwo}%
\providecommand \translation [1]{[#1]}%
\providecommand \BibitemOpen [0]{}%
\providecommand \bibitemStop [0]{}%
\providecommand \bibitemNoStop [0]{.\EOS\space}%
\providecommand \EOS [0]{\spacefactor3000\relax}%
\providecommand \BibitemShut  [1]{\csname bibitem#1\endcsname}%
\let\auto@bib@innerbib\@empty
\bibitem [{\citenamefont {Calegari}\ and\ \citenamefont {Martin}(2023)}]{Calegari2023_Open}%
  \BibitemOpen
  \bibfield  {author} {\bibinfo {author} {\bibfnamefont {F.}~\bibnamefont {Calegari}}\ and\ \bibinfo {author} {\bibfnamefont {F.}~\bibnamefont {Martin}},\ }\bibfield  {title} {\bibinfo {title} {Open questions in attochemistry},\ }\href {https://doi.org/10.1038/s42004-023-00989-0} {\bibfield  {journal} {\bibinfo  {journal} {Commun Chem}\ }\textbf {\bibinfo {volume} {6}},\ \bibinfo {pages} {1} (\bibinfo {year} {2023})}\BibitemShut {NoStop}%
\bibitem [{\citenamefont {Ivanov}(2021)}]{Ivanov2021_Concluding}%
  \BibitemOpen
  \bibfield  {author} {\bibinfo {author} {\bibfnamefont {M.}~\bibnamefont {Ivanov}},\ }\bibfield  {title} {\bibinfo {title} {Concluding remarks: {{The}} age of molecular movies},\ }\href {https://doi.org/10.1039/D1FD90033A} {\bibfield  {journal} {\bibinfo  {journal} {Faraday Discuss.}\ }\textbf {\bibinfo {volume} {228}},\ \bibinfo {pages} {622} (\bibinfo {year} {2021})}\BibitemShut {NoStop}%
\bibitem [{\citenamefont {Cederbaum}\ and\ \citenamefont {Zobeley}(1999)}]{Cederbaum1999_Ultrafast}%
  \BibitemOpen
  \bibfield  {author} {\bibinfo {author} {\bibfnamefont {L.~S.}\ \bibnamefont {Cederbaum}}\ and\ \bibinfo {author} {\bibfnamefont {J.}~\bibnamefont {Zobeley}},\ }\bibfield  {title} {\bibinfo {title} {Ultrafast charge migration by electron correlation},\ }\href {https://doi.org/10.1016/S0009-2614(99)00508-4} {\bibfield  {journal} {\bibinfo  {journal} {Chem. Phys. Lett.}\ }\textbf {\bibinfo {volume} {307}},\ \bibinfo {pages} {205} (\bibinfo {year} {1999})}\BibitemShut {NoStop}%
\bibitem [{\citenamefont {Remacle}\ and\ \citenamefont {Levine}(2006)}]{Remacle2006_electronic}%
  \BibitemOpen
  \bibfield  {author} {\bibinfo {author} {\bibfnamefont {F.}~\bibnamefont {Remacle}}\ and\ \bibinfo {author} {\bibfnamefont {R.~D.}\ \bibnamefont {Levine}},\ }\bibfield  {title} {\bibinfo {title} {An electronic time scale in chemistry},\ }\href {https://doi.org/10.1073/pnas.0601855103} {\bibfield  {journal} {\bibinfo  {journal} {Proc. Natl. Acad. Sci.}\ }\textbf {\bibinfo {volume} {103}},\ \bibinfo {pages} {6793} (\bibinfo {year} {2006})}\BibitemShut {NoStop}%
\bibitem [{\citenamefont {Arnold}\ \emph {et~al.}(2017)\citenamefont {Arnold}, \citenamefont {Vendrell},\ and\ \citenamefont {Santra}}]{Arnold2017_Electronic}%
  \BibitemOpen
  \bibfield  {author} {\bibinfo {author} {\bibfnamefont {C.}~\bibnamefont {Arnold}}, \bibinfo {author} {\bibfnamefont {O.}~\bibnamefont {Vendrell}},\ and\ \bibinfo {author} {\bibfnamefont {R.}~\bibnamefont {Santra}},\ }\bibfield  {title} {\bibinfo {title} {Electronic decoherence following photoionization: {{Full}} quantum-dynamical treatment of the influence of nuclear motion},\ }\href {https://doi.org/10.1103/PhysRevA.95.033425} {\bibfield  {journal} {\bibinfo  {journal} {Phys. Rev. A}\ }\textbf {\bibinfo {volume} {95}},\ \bibinfo {pages} {033425} (\bibinfo {year} {2017})}\BibitemShut {NoStop}%
\bibitem [{\citenamefont {Remacle}\ \emph {et~al.}(1998)\citenamefont {Remacle}, \citenamefont {Levine},\ and\ \citenamefont {Ratner}}]{Remacle1998_Charge}%
  \BibitemOpen
  \bibfield  {author} {\bibinfo {author} {\bibfnamefont {F.}~\bibnamefont {Remacle}}, \bibinfo {author} {\bibfnamefont {{\relax RD}.}~\bibnamefont {Levine}},\ and\ \bibinfo {author} {\bibfnamefont {{\relax MA}.}~\bibnamefont {Ratner}},\ }\bibfield  {title} {\bibinfo {title} {Charge directed reactivity:: A simple electronic model, exhibiting site selectivity, for the dissociation of ions},\ }\href {https://doi.org/10.1016/S0009-2614(97)01314-6} {\bibfield  {journal} {\bibinfo  {journal} {Chem. Phys. Lett.}\ }\textbf {\bibinfo {volume} {285}},\ \bibinfo {pages} {25} (\bibinfo {year} {1998})}\BibitemShut {NoStop}%
\bibitem [{\citenamefont {L{\'e}pine}\ \emph {et~al.}(2014)\citenamefont {L{\'e}pine}, \citenamefont {Ivanov},\ and\ \citenamefont {Vrakking}}]{Lepine2014_Attosecond}%
  \BibitemOpen
  \bibfield  {author} {\bibinfo {author} {\bibfnamefont {F.}~\bibnamefont {L{\'e}pine}}, \bibinfo {author} {\bibfnamefont {M.~Y.}\ \bibnamefont {Ivanov}},\ and\ \bibinfo {author} {\bibfnamefont {M.~J.}\ \bibnamefont {Vrakking}},\ }\bibfield  {title} {\bibinfo {title} {Attosecond molecular dynamics: Fact or fiction?},\ }\href {https://doi.org/10.1038/nphoton.2014.25} {\bibfield  {journal} {\bibinfo  {journal} {Nat. Photonics}\ }\textbf {\bibinfo {volume} {8}},\ \bibinfo {pages} {195} (\bibinfo {year} {2014})}\BibitemShut {NoStop}%
\bibitem [{\citenamefont {Pianowski}(2022)}]{Pianowski2022_Molecular}%
  \BibitemOpen
  \bibfield  {author} {\bibinfo {author} {\bibfnamefont {Z.~L.}\ \bibnamefont {Pianowski}},\ }\href@noop {} {\emph {\bibinfo {title} {Molecular Photoswitches: {{Chemistry}}, Properties, and Applications Vols 1 and 2}}}\ (\bibinfo  {publisher} {Wiley},\ \bibinfo {year} {2022})\BibitemShut {NoStop}%
\bibitem [{\citenamefont {Schoenlein}\ \emph {et~al.}(1991)\citenamefont {Schoenlein}, \citenamefont {Peteanu}, \citenamefont {Mathies},\ and\ \citenamefont {Shank}}]{Schoenlein1991_first}%
  \BibitemOpen
  \bibfield  {author} {\bibinfo {author} {\bibfnamefont {R.~W.}\ \bibnamefont {Schoenlein}}, \bibinfo {author} {\bibfnamefont {L.~A.}\ \bibnamefont {Peteanu}}, \bibinfo {author} {\bibfnamefont {R.~A.}\ \bibnamefont {Mathies}},\ and\ \bibinfo {author} {\bibfnamefont {C.~V.}\ \bibnamefont {Shank}},\ }\bibfield  {title} {\bibinfo {title} {The first step in vision: {{Femtosecond}} isomerization of rhodopsin},\ }\href {https://doi.org/10.1126/science.1925597} {\bibfield  {journal} {\bibinfo  {journal} {Science}\ }\textbf {\bibinfo {volume} {254}},\ \bibinfo {pages} {412} (\bibinfo {year} {1991})}\BibitemShut {NoStop}%
\bibitem [{\citenamefont {Polli}\ \emph {et~al.}(2010)\citenamefont {Polli}, \citenamefont {Alto{\`e}}, \citenamefont {Weingart}, \citenamefont {Spillane}, \citenamefont {Manzoni}, \citenamefont {Brida}, \citenamefont {Tomasello}, \citenamefont {Orlandi}, \citenamefont {Kukura}, \citenamefont {Mathies}, \citenamefont {Garavelli},\ and\ \citenamefont {Cerullo}}]{Polli2010_Conical}%
  \BibitemOpen
  \bibfield  {author} {\bibinfo {author} {\bibfnamefont {D.}~\bibnamefont {Polli}}, \bibinfo {author} {\bibfnamefont {P.}~\bibnamefont {Alto{\`e}}}, \bibinfo {author} {\bibfnamefont {O.}~\bibnamefont {Weingart}}, \bibinfo {author} {\bibfnamefont {K.~M.}\ \bibnamefont {Spillane}}, \bibinfo {author} {\bibfnamefont {C.}~\bibnamefont {Manzoni}}, \bibinfo {author} {\bibfnamefont {D.}~\bibnamefont {Brida}}, \bibinfo {author} {\bibfnamefont {G.}~\bibnamefont {Tomasello}}, \bibinfo {author} {\bibfnamefont {G.}~\bibnamefont {Orlandi}}, \bibinfo {author} {\bibfnamefont {P.}~\bibnamefont {Kukura}}, \bibinfo {author} {\bibfnamefont {R.~A.}\ \bibnamefont {Mathies}}, \bibinfo {author} {\bibfnamefont {M.}~\bibnamefont {Garavelli}},\ and\ \bibinfo {author} {\bibfnamefont {G.}~\bibnamefont {Cerullo}},\ }\bibfield  {title} {\bibinfo {title} {Conical intersection dynamics of the primary photoisomerization event in vision},\ }\href {https://doi.org/10.1038/nature09346} {\bibfield  {journal} {\bibinfo  {journal} {Nature}\
  }\textbf {\bibinfo {volume} {467}},\ \bibinfo {pages} {440} (\bibinfo {year} {2010})}\BibitemShut {NoStop}%
\bibitem [{\citenamefont {Herek}\ \emph {et~al.}(2002)\citenamefont {Herek}, \citenamefont {Wohlleben}, \citenamefont {Cogdell}, \citenamefont {Zeidler},\ and\ \citenamefont {Motzkus}}]{Herek2002_Quantum}%
  \BibitemOpen
  \bibfield  {author} {\bibinfo {author} {\bibfnamefont {J.~L.}\ \bibnamefont {Herek}}, \bibinfo {author} {\bibfnamefont {W.}~\bibnamefont {Wohlleben}}, \bibinfo {author} {\bibfnamefont {R.~J.}\ \bibnamefont {Cogdell}}, \bibinfo {author} {\bibfnamefont {D.}~\bibnamefont {Zeidler}},\ and\ \bibinfo {author} {\bibfnamefont {M.}~\bibnamefont {Motzkus}},\ }\bibfield  {title} {\bibinfo {title} {Quantum control of energy flow in light harvesting},\ }\href {https://doi.org/10.1038/417533a} {\bibfield  {journal} {\bibinfo  {journal} {Nature}\ }\textbf {\bibinfo {volume} {417}},\ \bibinfo {pages} {533} (\bibinfo {year} {2002})}\BibitemShut {NoStop}%
\bibitem [{\citenamefont {Wang}\ \emph {et~al.}(2019)\citenamefont {Wang}, \citenamefont {Allodi},\ and\ \citenamefont {Engel}}]{Wang2019_Quantum}%
  \BibitemOpen
  \bibfield  {author} {\bibinfo {author} {\bibfnamefont {L.}~\bibnamefont {Wang}}, \bibinfo {author} {\bibfnamefont {M.~A.}\ \bibnamefont {Allodi}},\ and\ \bibinfo {author} {\bibfnamefont {G.~S.}\ \bibnamefont {Engel}},\ }\bibfield  {title} {\bibinfo {title} {Quantum coherences reveal excited-state dynamics in biophysical systems},\ }\href {https://doi.org/10.1038/s41570-019-0109-z} {\bibfield  {journal} {\bibinfo  {journal} {Nat. Rev. Chem.}\ }\textbf {\bibinfo {volume} {3}},\ \bibinfo {pages} {477} (\bibinfo {year} {2019})}\BibitemShut {NoStop}%
\bibitem [{\citenamefont {Cao}\ \emph {et~al.}(2020)\citenamefont {Cao}, \citenamefont {Cogdell}, \citenamefont {Coker}, \citenamefont {Duan}, \citenamefont {Hauer}, \citenamefont {Kleinekath{\"o}fer}, \citenamefont {Jansen}, \citenamefont {Man{\v c}al}, \citenamefont {Miller}, \citenamefont {Ogilvie}, \citenamefont {Prokhorenko}, \citenamefont {Renger}, \citenamefont {Tan}, \citenamefont {Tempelaar}, \citenamefont {Thorwart}, \citenamefont {Thyrhaug}, \citenamefont {Westenhoff},\ and\ \citenamefont {Zigmantas}}]{Cao2020_Quantum}%
  \BibitemOpen
  \bibfield  {author} {\bibinfo {author} {\bibfnamefont {J.}~\bibnamefont {Cao}}, \bibinfo {author} {\bibfnamefont {R.~J.}\ \bibnamefont {Cogdell}}, \bibinfo {author} {\bibfnamefont {D.~F.}\ \bibnamefont {Coker}}, \bibinfo {author} {\bibfnamefont {H.-G.}\ \bibnamefont {Duan}}, \bibinfo {author} {\bibfnamefont {J.}~\bibnamefont {Hauer}}, \bibinfo {author} {\bibfnamefont {U.}~\bibnamefont {Kleinekath{\"o}fer}}, \bibinfo {author} {\bibfnamefont {T.~L.~C.}\ \bibnamefont {Jansen}}, \bibinfo {author} {\bibfnamefont {T.}~\bibnamefont {Man{\v c}al}}, \bibinfo {author} {\bibfnamefont {R.~J.~D.}\ \bibnamefont {Miller}}, \bibinfo {author} {\bibfnamefont {J.~P.}\ \bibnamefont {Ogilvie}}, \bibinfo {author} {\bibfnamefont {V.~I.}\ \bibnamefont {Prokhorenko}}, \bibinfo {author} {\bibfnamefont {T.}~\bibnamefont {Renger}}, \bibinfo {author} {\bibfnamefont {H.-S.}\ \bibnamefont {Tan}}, \bibinfo {author} {\bibfnamefont {R.}~\bibnamefont {Tempelaar}}, \bibinfo {author} {\bibfnamefont {M.}~\bibnamefont {Thorwart}}, \bibinfo
  {author} {\bibfnamefont {E.}~\bibnamefont {Thyrhaug}}, \bibinfo {author} {\bibfnamefont {S.}~\bibnamefont {Westenhoff}},\ and\ \bibinfo {author} {\bibfnamefont {D.}~\bibnamefont {Zigmantas}},\ }\bibfield  {title} {\bibinfo {title} {Quantum biology revisited},\ }\href {https://doi.org/10.1126/sciadv.aaz4888} {\bibfield  {journal} {\bibinfo  {journal} {Sci. Adv.}\ }\textbf {\bibinfo {volume} {6}},\ \bibinfo {pages} {eaaz4888} (\bibinfo {year} {2020})}\BibitemShut {NoStop}%
\bibitem [{\citenamefont {Lambert}\ \emph {et~al.}(2013)\citenamefont {Lambert}, \citenamefont {Chen}, \citenamefont {Cheng}, \citenamefont {Li}, \citenamefont {Chen},\ and\ \citenamefont {Nori}}]{Lambert2013_Quantum}%
  \BibitemOpen
  \bibfield  {author} {\bibinfo {author} {\bibfnamefont {N.}~\bibnamefont {Lambert}}, \bibinfo {author} {\bibfnamefont {Y.-N.}\ \bibnamefont {Chen}}, \bibinfo {author} {\bibfnamefont {Y.-C.}\ \bibnamefont {Cheng}}, \bibinfo {author} {\bibfnamefont {C.-M.}\ \bibnamefont {Li}}, \bibinfo {author} {\bibfnamefont {G.-Y.}\ \bibnamefont {Chen}},\ and\ \bibinfo {author} {\bibfnamefont {F.}~\bibnamefont {Nori}},\ }\bibfield  {title} {\bibinfo {title} {Quantum biology},\ }\href {https://doi.org/10.1038/nphys2474} {\bibfield  {journal} {\bibinfo  {journal} {Nat. Phys.}\ }\textbf {\bibinfo {volume} {9}},\ \bibinfo {pages} {10} (\bibinfo {year} {2013})}\BibitemShut {NoStop}%
\bibitem [{\citenamefont {Wang}\ \emph {et~al.}(2022)\citenamefont {Wang}, \citenamefont {Xia},\ and\ \citenamefont {Ho}}]{Wang2022_Atomicscale}%
  \BibitemOpen
  \bibfield  {author} {\bibinfo {author} {\bibfnamefont {L.}~\bibnamefont {Wang}}, \bibinfo {author} {\bibfnamefont {Y.}~\bibnamefont {Xia}},\ and\ \bibinfo {author} {\bibfnamefont {W.}~\bibnamefont {Ho}},\ }\bibfield  {title} {\bibinfo {title} {Atomic-scale quantum sensing based on the ultrafast coherence of an {{H}}{$_2$} molecule in an {{STM}} cavity},\ }\href {https://doi.org/10.1126/science.abn9220} {\bibfield  {journal} {\bibinfo  {journal} {Science}\ }\textbf {\bibinfo {volume} {376}},\ \bibinfo {pages} {401} (\bibinfo {year} {2022})}\BibitemShut {NoStop}%
\bibitem [{\citenamefont {Wasielewski}\ \emph {et~al.}(2020)\citenamefont {Wasielewski}, \citenamefont {Forbes}, \citenamefont {Frank}, \citenamefont {Kowalski}, \citenamefont {Scholes}, \citenamefont {{Yuen-Zhou}}, \citenamefont {Baldo}, \citenamefont {Freedman}, \citenamefont {Goldsmith}, \citenamefont {Goodson~III} \emph {et~al.}}]{Wasielewski2020_Exploiting}%
  \BibitemOpen
  \bibfield  {author} {\bibinfo {author} {\bibfnamefont {M.~R.}\ \bibnamefont {Wasielewski}}, \bibinfo {author} {\bibfnamefont {M.~D.}\ \bibnamefont {Forbes}}, \bibinfo {author} {\bibfnamefont {N.~L.}\ \bibnamefont {Frank}}, \bibinfo {author} {\bibfnamefont {K.}~\bibnamefont {Kowalski}}, \bibinfo {author} {\bibfnamefont {G.~D.}\ \bibnamefont {Scholes}}, \bibinfo {author} {\bibfnamefont {J.}~\bibnamefont {{Yuen-Zhou}}}, \bibinfo {author} {\bibfnamefont {M.~A.}\ \bibnamefont {Baldo}}, \bibinfo {author} {\bibfnamefont {D.~E.}\ \bibnamefont {Freedman}}, \bibinfo {author} {\bibfnamefont {R.~H.}\ \bibnamefont {Goldsmith}}, \bibinfo {author} {\bibfnamefont {T.}~\bibnamefont {Goodson~III}}, \emph {et~al.},\ }\bibfield  {title} {\bibinfo {title} {Exploiting chemistry and molecular systems for quantum information science},\ }\href {https://doi.org/10.1038/s41570-020-0200-5} {\bibfield  {journal} {\bibinfo  {journal} {Nat. Rev. Chem.}\ }\textbf {\bibinfo {volume} {4}},\ \bibinfo {pages} {490} (\bibinfo {year}
  {2020})}\BibitemShut {NoStop}%
\bibitem [{\citenamefont {Smirnova}\ \emph {et~al.}(2009)\citenamefont {Smirnova}, \citenamefont {Mairesse}, \citenamefont {Patchkovskii}, \citenamefont {Dudovich}, \citenamefont {Villeneuve}, \citenamefont {Corkum},\ and\ \citenamefont {Ivanov}}]{Smirnova2009_High}%
  \BibitemOpen
  \bibfield  {author} {\bibinfo {author} {\bibfnamefont {O.}~\bibnamefont {Smirnova}}, \bibinfo {author} {\bibfnamefont {Y.}~\bibnamefont {Mairesse}}, \bibinfo {author} {\bibfnamefont {S.}~\bibnamefont {Patchkovskii}}, \bibinfo {author} {\bibfnamefont {N.}~\bibnamefont {Dudovich}}, \bibinfo {author} {\bibfnamefont {D.}~\bibnamefont {Villeneuve}}, \bibinfo {author} {\bibfnamefont {P.}~\bibnamefont {Corkum}},\ and\ \bibinfo {author} {\bibfnamefont {M.~Y.}\ \bibnamefont {Ivanov}},\ }\bibfield  {title} {\bibinfo {title} {High harmonic interferometry of multi-electron dynamics in molecules},\ }\href {https://doi.org/10.1038/nature08253} {\bibfield  {journal} {\bibinfo  {journal} {Nature}\ }\textbf {\bibinfo {volume} {460}},\ \bibinfo {pages} {972} (\bibinfo {year} {2009})}\BibitemShut {NoStop}%
\bibitem [{\citenamefont {Kraus}\ \emph {et~al.}(2015)\citenamefont {Kraus}, \citenamefont {Mignolet}, \citenamefont {Baykusheva}, \citenamefont {Rupenyan}, \citenamefont {Horn{\`y}}, \citenamefont {Penka}, \citenamefont {Grassi}, \citenamefont {Tolstikhin}, \citenamefont {Schneider}, \citenamefont {Jensen} \emph {et~al.}}]{Kraus2015_Measurement}%
  \BibitemOpen
  \bibfield  {author} {\bibinfo {author} {\bibfnamefont {P.~M.}\ \bibnamefont {Kraus}}, \bibinfo {author} {\bibfnamefont {B.}~\bibnamefont {Mignolet}}, \bibinfo {author} {\bibfnamefont {D.}~\bibnamefont {Baykusheva}}, \bibinfo {author} {\bibfnamefont {A.}~\bibnamefont {Rupenyan}}, \bibinfo {author} {\bibfnamefont {L.}~\bibnamefont {Horn{\`y}}}, \bibinfo {author} {\bibfnamefont {E.~F.}\ \bibnamefont {Penka}}, \bibinfo {author} {\bibfnamefont {G.}~\bibnamefont {Grassi}}, \bibinfo {author} {\bibfnamefont {O.~I.}\ \bibnamefont {Tolstikhin}}, \bibinfo {author} {\bibfnamefont {J.}~\bibnamefont {Schneider}}, \bibinfo {author} {\bibfnamefont {F.}~\bibnamefont {Jensen}}, \emph {et~al.},\ }\bibfield  {title} {\bibinfo {title} {Measurement and laser control of attosecond charge migration in ionized iodoacetylene},\ }\href {https://doi.org/10.1126/science.aab2160} {\bibfield  {journal} {\bibinfo  {journal} {Science}\ }\textbf {\bibinfo {volume} {350}},\ \bibinfo {pages} {790} (\bibinfo {year} {2015})}\BibitemShut
  {NoStop}%
\bibitem [{\citenamefont {He}\ \emph {et~al.}(2022)\citenamefont {He}, \citenamefont {Sun}, \citenamefont {Lan}, \citenamefont {He}, \citenamefont {Wang}, \citenamefont {Wang}, \citenamefont {Zhu}, \citenamefont {Li}, \citenamefont {Cao}, \citenamefont {Lu},\ and\ \citenamefont {Lin}}]{He2022_Filming}%
  \BibitemOpen
  \bibfield  {author} {\bibinfo {author} {\bibfnamefont {L.}~\bibnamefont {He}}, \bibinfo {author} {\bibfnamefont {S.}~\bibnamefont {Sun}}, \bibinfo {author} {\bibfnamefont {P.}~\bibnamefont {Lan}}, \bibinfo {author} {\bibfnamefont {Y.}~\bibnamefont {He}}, \bibinfo {author} {\bibfnamefont {B.}~\bibnamefont {Wang}}, \bibinfo {author} {\bibfnamefont {P.}~\bibnamefont {Wang}}, \bibinfo {author} {\bibfnamefont {X.}~\bibnamefont {Zhu}}, \bibinfo {author} {\bibfnamefont {L.}~\bibnamefont {Li}}, \bibinfo {author} {\bibfnamefont {W.}~\bibnamefont {Cao}}, \bibinfo {author} {\bibfnamefont {P.}~\bibnamefont {Lu}},\ and\ \bibinfo {author} {\bibfnamefont {C.~D.}\ \bibnamefont {Lin}},\ }\bibfield  {title} {\bibinfo {title} {Filming movies of attosecond charge migration in single molecules with high harmonic spectroscopy},\ }\href {https://doi.org/10.1038/s41467-022-32313-0} {\bibfield  {journal} {\bibinfo  {journal} {Nat. commun.}\ }\textbf {\bibinfo {volume} {13}},\ \bibinfo {pages} {1} (\bibinfo {year}
  {2022})}\BibitemShut {NoStop}%
\bibitem [{\citenamefont {He}\ \emph {et~al.}(2023)\citenamefont {He}, \citenamefont {He}, \citenamefont {Sun}, \citenamefont {Goetz}, \citenamefont {Le}, \citenamefont {Zhu}, \citenamefont {Lan}, \citenamefont {Lu},\ and\ \citenamefont {Lin}}]{He2023_Attosecond}%
  \BibitemOpen
  \bibfield  {author} {\bibinfo {author} {\bibfnamefont {L.}~\bibnamefont {He}}, \bibinfo {author} {\bibfnamefont {Y.}~\bibnamefont {He}}, \bibinfo {author} {\bibfnamefont {S.}~\bibnamefont {Sun}}, \bibinfo {author} {\bibfnamefont {E.}~\bibnamefont {Goetz}}, \bibinfo {author} {\bibfnamefont {A.-T.}\ \bibnamefont {Le}}, \bibinfo {author} {\bibfnamefont {X.}~\bibnamefont {Zhu}}, \bibinfo {author} {\bibfnamefont {P.}~\bibnamefont {Lan}}, \bibinfo {author} {\bibfnamefont {P.}~\bibnamefont {Lu}},\ and\ \bibinfo {author} {\bibfnamefont {C.-D.}\ \bibnamefont {Lin}},\ }\bibfield  {title} {\bibinfo {title} {Attosecond probing and control of charge migration in carbon-chain molecule},\ }\href {https://doi.org/10.1117/1.AP.5.5.056001} {\bibfield  {journal} {\bibinfo  {journal} {Adv. Photon.}\ }\textbf {\bibinfo {volume} {5}},\ \bibinfo {pages} {056001} (\bibinfo {year} {2023})}\BibitemShut {NoStop}%
\bibitem [{\citenamefont {Calegari}\ \emph {et~al.}(2014)\citenamefont {Calegari}, \citenamefont {Ayuso}, \citenamefont {Trabattoni}, \citenamefont {Belshaw}, \citenamefont {De~Camillis}, \citenamefont {Anumula}, \citenamefont {Frassetto}, \citenamefont {Poletto}, \citenamefont {Palacios}, \citenamefont {Decleva}, \citenamefont {Greenwood}, \citenamefont {Martin},\ and\ \citenamefont {Nisoli}}]{Calegari2014_Ultrafast}%
  \BibitemOpen
  \bibfield  {author} {\bibinfo {author} {\bibfnamefont {F.}~\bibnamefont {Calegari}}, \bibinfo {author} {\bibfnamefont {D.}~\bibnamefont {Ayuso}}, \bibinfo {author} {\bibfnamefont {A.}~\bibnamefont {Trabattoni}}, \bibinfo {author} {\bibfnamefont {L.}~\bibnamefont {Belshaw}}, \bibinfo {author} {\bibfnamefont {S.}~\bibnamefont {De~Camillis}}, \bibinfo {author} {\bibfnamefont {S.}~\bibnamefont {Anumula}}, \bibinfo {author} {\bibfnamefont {F.}~\bibnamefont {Frassetto}}, \bibinfo {author} {\bibfnamefont {L.}~\bibnamefont {Poletto}}, \bibinfo {author} {\bibfnamefont {A.}~\bibnamefont {Palacios}}, \bibinfo {author} {\bibfnamefont {P.}~\bibnamefont {Decleva}}, \bibinfo {author} {\bibfnamefont {J.~B.}\ \bibnamefont {Greenwood}}, \bibinfo {author} {\bibfnamefont {F.}~\bibnamefont {Martin}},\ and\ \bibinfo {author} {\bibfnamefont {M.}~\bibnamefont {Nisoli}},\ }\bibfield  {title} {\bibinfo {title} {Ultrafast electron dynamics in phenylalanine initiated by attosecond pulses},\ }\href
  {https://doi.org/10.1126/science.1254061} {\bibfield  {journal} {\bibinfo  {journal} {Science}\ }\textbf {\bibinfo {volume} {346}},\ \bibinfo {pages} {336} (\bibinfo {year} {2014})}\BibitemShut {NoStop}%
\bibitem [{\citenamefont {{Lara-Astiaso}}\ \emph {et~al.}(2018)\citenamefont {{Lara-Astiaso}}, \citenamefont {Galli}, \citenamefont {Trabattoni}, \citenamefont {Palacios}, \citenamefont {Ayuso}, \citenamefont {Frassetto}, \citenamefont {Poletto}, \citenamefont {De~Camillis}, \citenamefont {Greenwood}, \citenamefont {Decleva}, \citenamefont {Tavernelli}, \citenamefont {Calegari}, \citenamefont {Nisoli},\ and\ \citenamefont {Martin}}]{Lara-Astiaso2018_Attosecond}%
  \BibitemOpen
  \bibfield  {author} {\bibinfo {author} {\bibfnamefont {M.}~\bibnamefont {{Lara-Astiaso}}}, \bibinfo {author} {\bibfnamefont {M.}~\bibnamefont {Galli}}, \bibinfo {author} {\bibfnamefont {A.}~\bibnamefont {Trabattoni}}, \bibinfo {author} {\bibfnamefont {A.}~\bibnamefont {Palacios}}, \bibinfo {author} {\bibfnamefont {D.}~\bibnamefont {Ayuso}}, \bibinfo {author} {\bibfnamefont {F.}~\bibnamefont {Frassetto}}, \bibinfo {author} {\bibfnamefont {L.}~\bibnamefont {Poletto}}, \bibinfo {author} {\bibfnamefont {S.}~\bibnamefont {De~Camillis}}, \bibinfo {author} {\bibfnamefont {J.}~\bibnamefont {Greenwood}}, \bibinfo {author} {\bibfnamefont {P.}~\bibnamefont {Decleva}}, \bibinfo {author} {\bibfnamefont {I.}~\bibnamefont {Tavernelli}}, \bibinfo {author} {\bibfnamefont {F.}~\bibnamefont {Calegari}}, \bibinfo {author} {\bibfnamefont {M.}~\bibnamefont {Nisoli}},\ and\ \bibinfo {author} {\bibfnamefont {F.}~\bibnamefont {Martin}},\ }\bibfield  {title} {\bibinfo {title} {Attosecond pump--probe spectroscopy of charge dynamics
  in tryptophan},\ }\href {https://doi.org/10.1021/acs.jpclett.8b01786} {\bibfield  {journal} {\bibinfo  {journal} {J. Phys. Chem. Lett.}\ }\textbf {\bibinfo {volume} {9}},\ \bibinfo {pages} {4570} (\bibinfo {year} {2018})}\BibitemShut {NoStop}%
\bibitem [{\citenamefont {Herv{\'e}}\ \emph {et~al.}(2021)\citenamefont {Herv{\'e}}, \citenamefont {Despr{\'e}}, \citenamefont {Castellanos~Nash}, \citenamefont {Loriot}, \citenamefont {Boyer}, \citenamefont {Scognamiglio}, \citenamefont {Karras}, \citenamefont {Br{\'e}dy}, \citenamefont {Constant}, \citenamefont {Tielens}, \citenamefont {Kuleff},\ and\ \citenamefont {L{\'e}pine}}]{Herve2021_Ultrafast}%
  \BibitemOpen
  \bibfield  {author} {\bibinfo {author} {\bibfnamefont {M.}~\bibnamefont {Herv{\'e}}}, \bibinfo {author} {\bibfnamefont {V.}~\bibnamefont {Despr{\'e}}}, \bibinfo {author} {\bibfnamefont {P.}~\bibnamefont {Castellanos~Nash}}, \bibinfo {author} {\bibfnamefont {V.}~\bibnamefont {Loriot}}, \bibinfo {author} {\bibfnamefont {A.}~\bibnamefont {Boyer}}, \bibinfo {author} {\bibfnamefont {A.}~\bibnamefont {Scognamiglio}}, \bibinfo {author} {\bibfnamefont {G.}~\bibnamefont {Karras}}, \bibinfo {author} {\bibfnamefont {R.}~\bibnamefont {Br{\'e}dy}}, \bibinfo {author} {\bibfnamefont {E.}~\bibnamefont {Constant}}, \bibinfo {author} {\bibfnamefont {A.~G. G.~M.}\ \bibnamefont {Tielens}}, \bibinfo {author} {\bibfnamefont {A.~I.}\ \bibnamefont {Kuleff}},\ and\ \bibinfo {author} {\bibfnamefont {F.}~\bibnamefont {L{\'e}pine}},\ }\bibfield  {title} {\bibinfo {title} {Ultrafast dynamics of correlation bands following {{XUV}} molecular photoionization},\ }\href {https://doi.org/10.1038/s41567-020-01073-3} {\bibfield  {journal}
  {\bibinfo  {journal} {Nat. Phys.}\ }\textbf {\bibinfo {volume} {17}},\ \bibinfo {pages} {327} (\bibinfo {year} {2021})}\BibitemShut {NoStop}%
\bibitem [{\citenamefont {Okino}\ \emph {et~al.}(2015)\citenamefont {Okino}, \citenamefont {Furukawa}, \citenamefont {Nabekawa}, \citenamefont {Miyabe}, \citenamefont {Amani~Eilanlou}, \citenamefont {Takahashi}, \citenamefont {Yamanouchi},\ and\ \citenamefont {Midorikawa}}]{Okino2015_Direct}%
  \BibitemOpen
  \bibfield  {author} {\bibinfo {author} {\bibfnamefont {T.}~\bibnamefont {Okino}}, \bibinfo {author} {\bibfnamefont {Y.}~\bibnamefont {Furukawa}}, \bibinfo {author} {\bibfnamefont {Y.}~\bibnamefont {Nabekawa}}, \bibinfo {author} {\bibfnamefont {S.}~\bibnamefont {Miyabe}}, \bibinfo {author} {\bibfnamefont {A.}~\bibnamefont {Amani~Eilanlou}}, \bibinfo {author} {\bibfnamefont {E.~J.}\ \bibnamefont {Takahashi}}, \bibinfo {author} {\bibfnamefont {K.}~\bibnamefont {Yamanouchi}},\ and\ \bibinfo {author} {\bibfnamefont {K.}~\bibnamefont {Midorikawa}},\ }\bibfield  {title} {\bibinfo {title} {Direct observation of an attosecond electron wave packet in a nitrogen molecule},\ }\href {https://doi.org/10.1126/sciadv.1500356} {\bibfield  {journal} {\bibinfo  {journal} {Sci. Adv.}\ }\textbf {\bibinfo {volume} {1}},\ \bibinfo {pages} {e1500356} (\bibinfo {year} {2015})}\BibitemShut {NoStop}%
\bibitem [{\citenamefont {Fukahori}\ \emph {et~al.}(2020)\citenamefont {Fukahori}, \citenamefont {Matsubara}, \citenamefont {Nabekawa}, \citenamefont {Yamanouchi},\ and\ \citenamefont {Midorikawa}}]{Fukahori2020_Ultrafast}%
  \BibitemOpen
  \bibfield  {author} {\bibinfo {author} {\bibfnamefont {S.}~\bibnamefont {Fukahori}}, \bibinfo {author} {\bibfnamefont {T.}~\bibnamefont {Matsubara}}, \bibinfo {author} {\bibfnamefont {Y.}~\bibnamefont {Nabekawa}}, \bibinfo {author} {\bibfnamefont {K.}~\bibnamefont {Yamanouchi}},\ and\ \bibinfo {author} {\bibfnamefont {K.}~\bibnamefont {Midorikawa}},\ }\bibfield  {title} {\bibinfo {title} {Ultrafast electron--nuclear wavepacket in generated and probed with attosecond pulse trains},\ }\href {https://doi.org/10.1088/1361-6455/ab94cc} {\bibfield  {journal} {\bibinfo  {journal} {J Phys B Mol Opt Phys}\ }\textbf {\bibinfo {volume} {53}},\ \bibinfo {pages} {164001} (\bibinfo {year} {2020})}\BibitemShut {NoStop}%
\bibitem [{\citenamefont {Barillot}\ \emph {et~al.}(2021)\citenamefont {Barillot}, \citenamefont {Alexander}, \citenamefont {Cooper}, \citenamefont {Driver}, \citenamefont {Garratt}, \citenamefont {Li}, \citenamefont {Al~Haddad}, \citenamefont {{Sanchez-Gonzalez}}, \citenamefont {Ag{\aa}ker}, \citenamefont {Arrell} \emph {et~al.}}]{Barillot2021_Correlationdriven}%
  \BibitemOpen
  \bibfield  {author} {\bibinfo {author} {\bibfnamefont {T.}~\bibnamefont {Barillot}}, \bibinfo {author} {\bibfnamefont {O.}~\bibnamefont {Alexander}}, \bibinfo {author} {\bibfnamefont {B.}~\bibnamefont {Cooper}}, \bibinfo {author} {\bibfnamefont {T.}~\bibnamefont {Driver}}, \bibinfo {author} {\bibfnamefont {D.}~\bibnamefont {Garratt}}, \bibinfo {author} {\bibfnamefont {S.}~\bibnamefont {Li}}, \bibinfo {author} {\bibfnamefont {A.}~\bibnamefont {Al~Haddad}}, \bibinfo {author} {\bibfnamefont {A.}~\bibnamefont {{Sanchez-Gonzalez}}}, \bibinfo {author} {\bibfnamefont {M.}~\bibnamefont {Ag{\aa}ker}}, \bibinfo {author} {\bibfnamefont {C.}~\bibnamefont {Arrell}}, \emph {et~al.},\ }\bibfield  {title} {\bibinfo {title} {Correlation-driven transient hole dynamics resolved in space and time in the isopropanol molecule},\ }\href {https://doi.org/10.1103/PhysRevX.11.031048} {\bibfield  {journal} {\bibinfo  {journal} {Phys. Rev. X}\ }\textbf {\bibinfo {volume} {11}},\ \bibinfo {pages} {031048} (\bibinfo {year}
  {2021})}\BibitemShut {NoStop}%
\bibitem [{\citenamefont {Schwickert}\ \emph {et~al.}(2022)\citenamefont {Schwickert}, \citenamefont {Ruberti}, \citenamefont {Koloren{\v c}}, \citenamefont {Usenko}, \citenamefont {Przystawik}, \citenamefont {Baev}, \citenamefont {Baev}, \citenamefont {Braune}, \citenamefont {Bocklage}, \citenamefont {Czwalinna} \emph {et~al.}}]{Schwickert2022_Electronic}%
  \BibitemOpen
  \bibfield  {author} {\bibinfo {author} {\bibfnamefont {D.}~\bibnamefont {Schwickert}}, \bibinfo {author} {\bibfnamefont {M.}~\bibnamefont {Ruberti}}, \bibinfo {author} {\bibfnamefont {P.}~\bibnamefont {Koloren{\v c}}}, \bibinfo {author} {\bibfnamefont {S.}~\bibnamefont {Usenko}}, \bibinfo {author} {\bibfnamefont {A.}~\bibnamefont {Przystawik}}, \bibinfo {author} {\bibfnamefont {K.}~\bibnamefont {Baev}}, \bibinfo {author} {\bibfnamefont {I.}~\bibnamefont {Baev}}, \bibinfo {author} {\bibfnamefont {M.}~\bibnamefont {Braune}}, \bibinfo {author} {\bibfnamefont {L.}~\bibnamefont {Bocklage}}, \bibinfo {author} {\bibfnamefont {M.~K.}\ \bibnamefont {Czwalinna}}, \emph {et~al.},\ }\bibfield  {title} {\bibinfo {title} {Electronic quantum coherence in glycine molecules probed with ultrashort x-ray pulses in real time},\ }\href {https://doi.org/10.1126/sciadv.abn6848} {\bibfield  {journal} {\bibinfo  {journal} {Sci. Adv.}\ }\textbf {\bibinfo {volume} {8}},\ \bibinfo {pages} {eabn6848} (\bibinfo {year}
  {2022})}\BibitemShut {NoStop}%
\bibitem [{\citenamefont {Kobayashi}\ \emph {et~al.}(2020{\natexlab{a}})\citenamefont {Kobayashi}, \citenamefont {Neumark},\ and\ \citenamefont {Leone}}]{Kobayashi2020_Attosecond}%
  \BibitemOpen
  \bibfield  {author} {\bibinfo {author} {\bibfnamefont {Y.}~\bibnamefont {Kobayashi}}, \bibinfo {author} {\bibfnamefont {D.~M.}\ \bibnamefont {Neumark}},\ and\ \bibinfo {author} {\bibfnamefont {S.~R.}\ \bibnamefont {Leone}},\ }\bibfield  {title} {\bibinfo {title} {Attosecond {{XUV}} probing of vibronic quantum superpositions in {{Br}}{$_{2}^{+}$}},\ }\href {https://doi.org/10.1103/PhysRevA.102.051102} {\bibfield  {journal} {\bibinfo  {journal} {Phys. Rev. A}\ }\textbf {\bibinfo {volume} {102}},\ \bibinfo {pages} {051102(R)} (\bibinfo {year} {2020}{\natexlab{a}})}\BibitemShut {NoStop}%
\bibitem [{\citenamefont {Kobayashi}\ \emph {et~al.}(2020{\natexlab{b}})\citenamefont {Kobayashi}, \citenamefont {Chang}, \citenamefont {Poullain}, \citenamefont {Scutelnic}, \citenamefont {Zeng}, \citenamefont {Neumark},\ and\ \citenamefont {Leone}}]{Kobayashi2020_Coherent}%
  \BibitemOpen
  \bibfield  {author} {\bibinfo {author} {\bibfnamefont {Y.}~\bibnamefont {Kobayashi}}, \bibinfo {author} {\bibfnamefont {K.~F.}\ \bibnamefont {Chang}}, \bibinfo {author} {\bibfnamefont {S.~M.}\ \bibnamefont {Poullain}}, \bibinfo {author} {\bibfnamefont {V.}~\bibnamefont {Scutelnic}}, \bibinfo {author} {\bibfnamefont {T.}~\bibnamefont {Zeng}}, \bibinfo {author} {\bibfnamefont {D.~M.}\ \bibnamefont {Neumark}},\ and\ \bibinfo {author} {\bibfnamefont {S.~R.}\ \bibnamefont {Leone}},\ }\bibfield  {title} {\bibinfo {title} {Coherent electronic-vibrational dynamics in deuterium bromide probed via attosecond transient-absorption spectroscopy},\ }\href {https://doi.org/10.1103/PhysRevA.101.063414} {\bibfield  {journal} {\bibinfo  {journal} {Phys. Rev. A}\ }\textbf {\bibinfo {volume} {101}},\ \bibinfo {pages} {063414} (\bibinfo {year} {2020}{\natexlab{b}})}\BibitemShut {NoStop}%
\bibitem [{\citenamefont {Matselyukh}\ \emph {et~al.}(2022)\citenamefont {Matselyukh}, \citenamefont {Despr{\'e}}, \citenamefont {Golubev}, \citenamefont {Kuleff},\ and\ \citenamefont {W{\"o}rner}}]{Matselyukh2022_Decoherence}%
  \BibitemOpen
  \bibfield  {author} {\bibinfo {author} {\bibfnamefont {D.~T.}\ \bibnamefont {Matselyukh}}, \bibinfo {author} {\bibfnamefont {V.}~\bibnamefont {Despr{\'e}}}, \bibinfo {author} {\bibfnamefont {N.~V.}\ \bibnamefont {Golubev}}, \bibinfo {author} {\bibfnamefont {A.~I.}\ \bibnamefont {Kuleff}},\ and\ \bibinfo {author} {\bibfnamefont {H.~J.}\ \bibnamefont {W{\"o}rner}},\ }\bibfield  {title} {\bibinfo {title} {Decoherence and revival in attosecond charge migration driven by non-adiabatic dynamics},\ }\href {https://doi.org/10.1038/s41567-022-01690-0} {\bibfield  {journal} {\bibinfo  {journal} {Nat. Phys.}\ }\textbf {\bibinfo {volume} {18}},\ \bibinfo {pages} {1206} (\bibinfo {year} {2022})}\BibitemShut {NoStop}%
\bibitem [{\citenamefont {Golubev}\ \emph {et~al.}(2021)\citenamefont {Golubev}, \citenamefont {Van{\'{\i}}{\v c}ek},\ and\ \citenamefont {Kuleff}}]{Golubev2021_Corevalence}%
  \BibitemOpen
  \bibfield  {author} {\bibinfo {author} {\bibfnamefont {N.~V.}\ \bibnamefont {Golubev}}, \bibinfo {author} {\bibfnamefont {J.}~\bibnamefont {Van{\'{\i}}{\v c}ek}},\ and\ \bibinfo {author} {\bibfnamefont {A.~I.}\ \bibnamefont {Kuleff}},\ }\bibfield  {title} {\bibinfo {title} {Core-valence attosecond transient absorption spectroscopy of polyatomic molecules},\ }\href {https://doi.org/10.1103/PhysRevLett.127.123001} {\bibfield  {journal} {\bibinfo  {journal} {Phys. Rev. Lett.}\ }\textbf {\bibinfo {volume} {127}},\ \bibinfo {pages} {123001} (\bibinfo {year} {2021})}\BibitemShut {NoStop}%
\bibitem [{\citenamefont {Kobayashi}\ \emph {et~al.}(2022)\citenamefont {Kobayashi}, \citenamefont {Neumark},\ and\ \citenamefont {Leone}}]{Kobayashi2022_Theoretical}%
  \BibitemOpen
  \bibfield  {author} {\bibinfo {author} {\bibfnamefont {Y.}~\bibnamefont {Kobayashi}}, \bibinfo {author} {\bibfnamefont {D.~M.}\ \bibnamefont {Neumark}},\ and\ \bibinfo {author} {\bibfnamefont {S.~R.}\ \bibnamefont {Leone}},\ }\bibfield  {title} {\bibinfo {title} {Theoretical analysis of the role of complex transition dipole phase in {{XUV}} transient-absorption probing of charge migration},\ }\href {https://doi.org/10.1364/OE.451129} {\bibfield  {journal} {\bibinfo  {journal} {Opt. Express}\ }\textbf {\bibinfo {volume} {30}},\ \bibinfo {pages} {5673} (\bibinfo {year} {2022})}\BibitemShut {NoStop}%
\bibitem [{\citenamefont {Yuen}\ and\ \citenamefont {Lin}(2024{\natexlab{a}})}]{Yuen2024_Rotation}%
  \BibitemOpen
  \bibfield  {author} {\bibinfo {author} {\bibfnamefont {C.-H.}\ \bibnamefont {Yuen}}\ and\ \bibinfo {author} {\bibfnamefont {C.-D.}\ \bibnamefont {Lin}},\ }\bibfield  {title} {\bibinfo {title} {Rotation in attosecond vibronic coherence spectroscopy for molecules},\ }\href {https://doi.org/10.1038/s42005-024-01607-8} {\bibfield  {journal} {\bibinfo  {journal} {Commun. Phys.}\ }\textbf {\bibinfo {volume} {7}},\ \bibinfo {pages} {115} (\bibinfo {year} {2024}{\natexlab{a}})}\BibitemShut {NoStop}%
\bibitem [{\citenamefont {Kaufman}\ \emph {et~al.}(2023)\citenamefont {Kaufman}, \citenamefont {Marquetand}, \citenamefont {Rozgonyi},\ and\ \citenamefont {Weinacht}}]{Kaufman2023_Longlived}%
  \BibitemOpen
  \bibfield  {author} {\bibinfo {author} {\bibfnamefont {B.}~\bibnamefont {Kaufman}}, \bibinfo {author} {\bibfnamefont {P.}~\bibnamefont {Marquetand}}, \bibinfo {author} {\bibfnamefont {T.}~\bibnamefont {Rozgonyi}},\ and\ \bibinfo {author} {\bibfnamefont {T.}~\bibnamefont {Weinacht}},\ }\bibfield  {title} {\bibinfo {title} {Long-lived electronic coherences in molecules},\ }\href {https://doi.org/10.1103/PhysRevLett.131.263202} {\bibfield  {journal} {\bibinfo  {journal} {Phys. Rev. Lett.}\ }\textbf {\bibinfo {volume} {131}},\ \bibinfo {pages} {263202} (\bibinfo {year} {2023})}\BibitemShut {NoStop}%
\bibitem [{\citenamefont {Yuen}\ and\ \citenamefont {Lin}(2024{\natexlab{b}})}]{Yuen2024_Probing}%
  \BibitemOpen
  \bibfield  {author} {\bibinfo {author} {\bibfnamefont {C.~H.}\ \bibnamefont {Yuen}}\ and\ \bibinfo {author} {\bibfnamefont {C.~D.}\ \bibnamefont {Lin}},\ }\bibfield  {title} {\bibinfo {title} {Probing vibronic coherence in charge migration in molecules using strong-field sequential double ionization},\ }\href {https://doi.org/10.1103/PhysRevA.109.L011101} {\bibfield  {journal} {\bibinfo  {journal} {Phys. Rev. A}\ }\textbf {\bibinfo {volume} {109}},\ \bibinfo {pages} {L011101} (\bibinfo {year} {2024}{\natexlab{b}})}\BibitemShut {NoStop}%
\bibitem [{\citenamefont {Yuen}\ and\ \citenamefont {Lin}(2023)}]{Yuen2023_Coherence}%
  \BibitemOpen
  \bibfield  {author} {\bibinfo {author} {\bibfnamefont {C.~H.}\ \bibnamefont {Yuen}}\ and\ \bibinfo {author} {\bibfnamefont {C.~D.}\ \bibnamefont {Lin}},\ }\bibfield  {title} {\bibinfo {title} {Coherence from multiorbital tunneling ionization of molecules},\ }\href {https://doi.org/10.1103/PhysRevA.108.023123} {\bibfield  {journal} {\bibinfo  {journal} {Phys. Rev. A}\ }\textbf {\bibinfo {volume} {108}},\ \bibinfo {pages} {023123} (\bibinfo {year} {2023})}\BibitemShut {NoStop}%
\bibitem [{\citenamefont {Yuen}\ and\ \citenamefont {Lin}(2022)}]{Yuen2022_Densitymatrix}%
  \BibitemOpen
  \bibfield  {author} {\bibinfo {author} {\bibfnamefont {C.~H.}\ \bibnamefont {Yuen}}\ and\ \bibinfo {author} {\bibfnamefont {C.~D.}\ \bibnamefont {Lin}},\ }\bibfield  {title} {\bibinfo {title} {Density-matrix approach for sequential dissociative double ionization of molecules},\ }\href {https://doi.org/10.1103/PhysRevA.106.023120} {\bibfield  {journal} {\bibinfo  {journal} {Phys. Rev. A}\ }\textbf {\bibinfo {volume} {106}},\ \bibinfo {pages} {023120} (\bibinfo {year} {2022})}\BibitemShut {NoStop}%
\bibitem [{\citenamefont {Jia}\ \emph {et~al.}(2024)\citenamefont {Jia}, \citenamefont {Yuen}, \citenamefont {Jing}, \citenamefont {Zhou}, \citenamefont {Lin},\ and\ \citenamefont {Zhao}}]{Jia2024_Improved}%
  \BibitemOpen
  \bibfield  {author} {\bibinfo {author} {\bibfnamefont {Y.-W.}\ \bibnamefont {Jia}}, \bibinfo {author} {\bibfnamefont {C.~H.}\ \bibnamefont {Yuen}}, \bibinfo {author} {\bibfnamefont {W.-Q.}\ \bibnamefont {Jing}}, \bibinfo {author} {\bibfnamefont {Z.-Y.}\ \bibnamefont {Zhou}}, \bibinfo {author} {\bibfnamefont {C.~D.}\ \bibnamefont {Lin}},\ and\ \bibinfo {author} {\bibfnamefont {S.-F.}\ \bibnamefont {Zhao}},\ }\bibfield  {title} {\bibinfo {title} {Improved model for the dissociative sequential double ionization of nitrogen molecules by an intense few-cycle infrared laser pulse},\ }\href {https://doi.org/10.1103/PhysRevA.110.023112} {\bibfield  {journal} {\bibinfo  {journal} {Phys. Rev. A}\ }\textbf {\bibinfo {volume} {110}},\ \bibinfo {pages} {023112} (\bibinfo {year} {2024})}\BibitemShut {NoStop}%
\bibitem [{\citenamefont {Langhoff}\ \emph {et~al.}(1987)\citenamefont {Langhoff}, \citenamefont {Bauschlicher~Jr},\ and\ \citenamefont {Partridge}}]{Langhoff1987_Theoretical}%
  \BibitemOpen
  \bibfield  {author} {\bibinfo {author} {\bibfnamefont {S.~R.}\ \bibnamefont {Langhoff}}, \bibinfo {author} {\bibfnamefont {C.~W.}\ \bibnamefont {Bauschlicher~Jr}},\ and\ \bibinfo {author} {\bibfnamefont {H.}~\bibnamefont {Partridge}},\ }\bibfield  {title} {\bibinfo {title} {Theoretical study of the {{N}}+ 2 {{Meinel}} system},\ }\href {https://doi.org/10.1063/1.452835} {\bibfield  {journal} {\bibinfo  {journal} {J. Chem. Phys.}\ }\textbf {\bibinfo {volume} {87}},\ \bibinfo {pages} {4716} (\bibinfo {year} {1987})}\BibitemShut {NoStop}%
\bibitem [{\citenamefont {Langhoff}\ and\ \citenamefont {Bauschlicher~Jr}(1988)}]{Langhoff1988_Theoretical}%
  \BibitemOpen
  \bibfield  {author} {\bibinfo {author} {\bibfnamefont {S.~R.}\ \bibnamefont {Langhoff}}\ and\ \bibinfo {author} {\bibfnamefont {C.~W.}\ \bibnamefont {Bauschlicher~Jr}},\ }\bibfield  {title} {\bibinfo {title} {Theoretical study of the first and second negative systems of {{N}}+ 2},\ }\href {https://doi.org/10.1063/1.454604} {\bibfield  {journal} {\bibinfo  {journal} {J. Chem. Phys.}\ }\textbf {\bibinfo {volume} {88}},\ \bibinfo {pages} {329} (\bibinfo {year} {1988})}\BibitemShut {NoStop}%
\bibitem [{\citenamefont {Fechner}\ \emph {et~al.}(2014)\citenamefont {Fechner}, \citenamefont {Camus}, \citenamefont {Ullrich}, \citenamefont {Pfeifer},\ and\ \citenamefont {Moshammer}}]{Fechner2014_StrongFielda}%
  \BibitemOpen
  \bibfield  {author} {\bibinfo {author} {\bibfnamefont {L.}~\bibnamefont {Fechner}}, \bibinfo {author} {\bibfnamefont {N.}~\bibnamefont {Camus}}, \bibinfo {author} {\bibfnamefont {J.}~\bibnamefont {Ullrich}}, \bibinfo {author} {\bibfnamefont {T.}~\bibnamefont {Pfeifer}},\ and\ \bibinfo {author} {\bibfnamefont {R.}~\bibnamefont {Moshammer}},\ }\bibfield  {title} {\bibinfo {title} {Strong-{{Field Tunneling}} from a {{Coherent Superposition}} of {{Electronic States}}},\ }\href {https://doi.org/10.1103/PhysRevLett.112.213001} {\bibfield  {journal} {\bibinfo  {journal} {Phys. Rev. Lett.}\ }\textbf {\bibinfo {volume} {112}},\ \bibinfo {pages} {213001} (\bibinfo {year} {2014})}\BibitemShut {NoStop}%
\bibitem [{\citenamefont {Robinson}\ \emph {et~al.}(2006)\citenamefont {Robinson}, \citenamefont {Haworth}, \citenamefont {Teng}, \citenamefont {Smith}, \citenamefont {Marangos},\ and\ \citenamefont {Tisch}}]{Robinson2006_generation}%
  \BibitemOpen
  \bibfield  {author} {\bibinfo {author} {\bibfnamefont {J.}~\bibnamefont {Robinson}}, \bibinfo {author} {\bibfnamefont {C.}~\bibnamefont {Haworth}}, \bibinfo {author} {\bibfnamefont {H.}~\bibnamefont {Teng}}, \bibinfo {author} {\bibfnamefont {R.}~\bibnamefont {Smith}}, \bibinfo {author} {\bibfnamefont {J.}~\bibnamefont {Marangos}},\ and\ \bibinfo {author} {\bibfnamefont {J.}~\bibnamefont {Tisch}},\ }\bibfield  {title} {\bibinfo {title} {The generation of intense, transform-limited laser pulses with tunable duration from 6 to 30~fs in a differentially pumped hollow fibre},\ }\href {https://doi.org/10.1007/s00340-006-2390-z} {\bibfield  {journal} {\bibinfo  {journal} {Appl. Phys. B}\ }\textbf {\bibinfo {volume} {85}},\ \bibinfo {pages} {525} (\bibinfo {year} {2006})}\BibitemShut {NoStop}%
\bibitem [{\citenamefont {Miranda}\ \emph {et~al.}(2012)\citenamefont {Miranda}, \citenamefont {Fordell}, \citenamefont {Arnold}, \citenamefont {L'Huillier},\ and\ \citenamefont {Crespo}}]{Miranda2012_Simultaneous}%
  \BibitemOpen
  \bibfield  {author} {\bibinfo {author} {\bibfnamefont {M.}~\bibnamefont {Miranda}}, \bibinfo {author} {\bibfnamefont {T.}~\bibnamefont {Fordell}}, \bibinfo {author} {\bibfnamefont {C.}~\bibnamefont {Arnold}}, \bibinfo {author} {\bibfnamefont {A.}~\bibnamefont {L'Huillier}},\ and\ \bibinfo {author} {\bibfnamefont {H.}~\bibnamefont {Crespo}},\ }\bibfield  {title} {\bibinfo {title} {Simultaneous compression and characterization of ultrashort laser pulses using chirped mirrors and glass wedges},\ }\href {https://doi.org/10.1364/OE.20.000688} {\bibfield  {journal} {\bibinfo  {journal} {Opt. Express, OE}\ }\textbf {\bibinfo {volume} {20}},\ \bibinfo {pages} {688} (\bibinfo {year} {2012})}\BibitemShut {NoStop}%
\bibitem [{\citenamefont {Sytcevich}\ \emph {et~al.}(2021)\citenamefont {Sytcevich}, \citenamefont {Guo}, \citenamefont {Mikaelsson}, \citenamefont {Vogelsang}, \citenamefont {Viotti}, \citenamefont {Alonso}, \citenamefont {Romero}, \citenamefont {Guerreiro}, \citenamefont {Sola}, \citenamefont {L'Huillier}, \citenamefont {Crespo}, \citenamefont {Miranda},\ and\ \citenamefont {Arnold}}]{Sytcevich2021_Characterizing}%
  \BibitemOpen
  \bibfield  {author} {\bibinfo {author} {\bibfnamefont {I.}~\bibnamefont {Sytcevich}}, \bibinfo {author} {\bibfnamefont {C.}~\bibnamefont {Guo}}, \bibinfo {author} {\bibfnamefont {S.}~\bibnamefont {Mikaelsson}}, \bibinfo {author} {\bibfnamefont {J.}~\bibnamefont {Vogelsang}}, \bibinfo {author} {\bibfnamefont {A.-L.}\ \bibnamefont {Viotti}}, \bibinfo {author} {\bibfnamefont {B.}~\bibnamefont {Alonso}}, \bibinfo {author} {\bibfnamefont {R.}~\bibnamefont {Romero}}, \bibinfo {author} {\bibfnamefont {P.~T.}\ \bibnamefont {Guerreiro}}, \bibinfo {author} {\bibfnamefont {{\'I}.~J.}\ \bibnamefont {Sola}}, \bibinfo {author} {\bibfnamefont {A.}~\bibnamefont {L'Huillier}}, \bibinfo {author} {\bibfnamefont {H.}~\bibnamefont {Crespo}}, \bibinfo {author} {\bibfnamefont {M.}~\bibnamefont {Miranda}},\ and\ \bibinfo {author} {\bibfnamefont {C.~L.}\ \bibnamefont {Arnold}},\ }\bibfield  {title} {\bibinfo {title} {Characterizing ultrashort laser pulses with second harmonic dispersion scans},\ }\href
  {https://doi.org/10.1364/JOSAB.412535} {\bibfield  {journal} {\bibinfo  {journal} {J. Opt. Soc. Am. B, JOSAB}\ }\textbf {\bibinfo {volume} {38}},\ \bibinfo {pages} {1546} (\bibinfo {year} {2021})}\BibitemShut {NoStop}%
\bibitem [{\citenamefont {Jagutzki}\ \emph {et~al.}(2002)\citenamefont {Jagutzki}, \citenamefont {Cerezo}, \citenamefont {Czasch}, \citenamefont {Dorner}, \citenamefont {Hattas}, \citenamefont {Huang}, \citenamefont {Mergel}, \citenamefont {Spillmann}, \citenamefont {{Ullmann-Pfleger}}, \citenamefont {Weber}, \citenamefont {{Schmidt-Bocking}},\ and\ \citenamefont {Smith}}]{Jagutzki2002_Multiple}%
  \BibitemOpen
  \bibfield  {author} {\bibinfo {author} {\bibfnamefont {O.}~\bibnamefont {Jagutzki}}, \bibinfo {author} {\bibfnamefont {A.}~\bibnamefont {Cerezo}}, \bibinfo {author} {\bibfnamefont {A.}~\bibnamefont {Czasch}}, \bibinfo {author} {\bibfnamefont {R.}~\bibnamefont {Dorner}}, \bibinfo {author} {\bibfnamefont {M.}~\bibnamefont {Hattas}}, \bibinfo {author} {\bibfnamefont {M.}~\bibnamefont {Huang}}, \bibinfo {author} {\bibfnamefont {V.}~\bibnamefont {Mergel}}, \bibinfo {author} {\bibfnamefont {U.}~\bibnamefont {Spillmann}}, \bibinfo {author} {\bibfnamefont {K.}~\bibnamefont {{Ullmann-Pfleger}}}, \bibinfo {author} {\bibfnamefont {T.}~\bibnamefont {Weber}}, \bibinfo {author} {\bibfnamefont {H.}~\bibnamefont {{Schmidt-Bocking}}},\ and\ \bibinfo {author} {\bibfnamefont {G.}~\bibnamefont {Smith}},\ }\bibfield  {title} {\bibinfo {title} {Multiple hit readout of a microchannel plate detector with a three-layer delay-line anode},\ }\href {https://doi.org/10.1109/TNS.2002.803889} {\bibfield  {journal} {\bibinfo  {journal}
  {IEEE Trans. Nucl. Sci.}\ }\textbf {\bibinfo {volume} {49}},\ \bibinfo {pages} {2477} (\bibinfo {year} {2002})}\BibitemShut {NoStop}%
\bibitem [{\citenamefont {Yuen}\ and\ \citenamefont {Lin}(2024{\natexlab{c}})}]{Yuen2024_Theory}%
  \BibitemOpen
  \bibfield  {author} {\bibinfo {author} {\bibfnamefont {C.~H.}\ \bibnamefont {Yuen}}\ and\ \bibinfo {author} {\bibfnamefont {C.~D.}\ \bibnamefont {Lin}},\ }\bibfield  {title} {\bibinfo {title} {Theory of strong-field sequential double ionization of polyatomic molecules},\ }\href {https://doi.org/10.1103/PhysRevA.109.033108} {\bibfield  {journal} {\bibinfo  {journal} {Phys. Rev. A}\ }\textbf {\bibinfo {volume} {109}},\ \bibinfo {pages} {033108} (\bibinfo {year} {2024}{\natexlab{c}})}\BibitemShut {NoStop}%
\bibitem [{\citenamefont {Tong}\ \emph {et~al.}(2002)\citenamefont {Tong}, \citenamefont {Zhao},\ and\ \citenamefont {Lin}}]{Tong2002_Theory}%
  \BibitemOpen
  \bibfield  {author} {\bibinfo {author} {\bibfnamefont {X.-M.}\ \bibnamefont {Tong}}, \bibinfo {author} {\bibfnamefont {Z.~X.}\ \bibnamefont {Zhao}},\ and\ \bibinfo {author} {\bibfnamefont {C.-D.}\ \bibnamefont {Lin}},\ }\bibfield  {title} {\bibinfo {title} {Theory of molecular tunneling ionization},\ }\href {https://doi.org/10.1103/PhysRevA.66.033402} {\bibfield  {journal} {\bibinfo  {journal} {Phys. Rev. A}\ }\textbf {\bibinfo {volume} {66}},\ \bibinfo {pages} {033402} (\bibinfo {year} {2002})}\BibitemShut {NoStop}%
\bibitem [{\citenamefont {Tolstikhin}\ \emph {et~al.}(2011)\citenamefont {Tolstikhin}, \citenamefont {Morishita},\ and\ \citenamefont {Madsen}}]{Tolstikhin2011_Theory}%
  \BibitemOpen
  \bibfield  {author} {\bibinfo {author} {\bibfnamefont {O.~I.}\ \bibnamefont {Tolstikhin}}, \bibinfo {author} {\bibfnamefont {T.}~\bibnamefont {Morishita}},\ and\ \bibinfo {author} {\bibfnamefont {L.~B.}\ \bibnamefont {Madsen}},\ }\bibfield  {title} {\bibinfo {title} {Theory of tunneling ionization of molecules: {{Weak-field}} asymptotics including dipole effects},\ }\href {https://doi.org/10.1103/PhysRevA.84.053423} {\bibfield  {journal} {\bibinfo  {journal} {Phys. Rev. A}\ }\textbf {\bibinfo {volume} {84}},\ \bibinfo {pages} {053423} (\bibinfo {year} {2011})}\BibitemShut {NoStop}%
\bibitem [{\citenamefont {Adrian N.~Pfeiffer}\ and\ \citenamefont {Leone}(2013)}]{AdrianN.Pfeiffer2013_Calculation}%
  \BibitemOpen
  \bibfield  {author} {\bibinfo {author} {\bibfnamefont {S.~G.~S.}\ \bibnamefont {Adrian N.~Pfeiffer}}\ and\ \bibinfo {author} {\bibfnamefont {S.~R.}\ \bibnamefont {Leone}},\ }\bibfield  {title} {\bibinfo {title} {Calculation of valence electron motion induced by sequential strong-field ionisation},\ }\href {https://doi.org/10.1080/00268976.2013.801527} {\bibfield  {journal} {\bibinfo  {journal} {Mol. Phys.}\ }\textbf {\bibinfo {volume} {111}},\ \bibinfo {pages} {2283} (\bibinfo {year} {2013})}\BibitemShut {NoStop}%
\bibitem [{\citenamefont {Tong}\ and\ \citenamefont {Lin}(2005)}]{Tong2005_Empirical}%
  \BibitemOpen
  \bibfield  {author} {\bibinfo {author} {\bibfnamefont {X.~M.}\ \bibnamefont {Tong}}\ and\ \bibinfo {author} {\bibfnamefont {C.~D.}\ \bibnamefont {Lin}},\ }\bibfield  {title} {\bibinfo {title} {Empirical formula for static field ionization rates of atoms and molecules by lasers in the barrier-suppression regime},\ }\href {https://doi.org/10.1088/0953-4075/38/15/001} {\bibfield  {journal} {\bibinfo  {journal} {J. Phys. B: At. Mol. Opt. Phys.}\ }\textbf {\bibinfo {volume} {38}},\ \bibinfo {pages} {2593} (\bibinfo {year} {2005})}\BibitemShut {NoStop}%
\bibitem [{\citenamefont {Bergues}\ \emph {et~al.}(2012)\citenamefont {Bergues}, \citenamefont {K{\"u}bel}, \citenamefont {Johnson}, \citenamefont {Fischer}, \citenamefont {Camus}, \citenamefont {Betsch}, \citenamefont {Herrwerth}, \citenamefont {Senftleben}, \citenamefont {Sayler}, \citenamefont {Rathje}, \citenamefont {Pfeifer}, \citenamefont {{Ben-Itzhak}}, \citenamefont {Jones}, \citenamefont {Paulus}, \citenamefont {Krausz}, \citenamefont {Moshammer}, \citenamefont {Ullrich},\ and\ \citenamefont {Kling}}]{Bergues2012_Attosecond}%
  \BibitemOpen
  \bibfield  {author} {\bibinfo {author} {\bibfnamefont {B.}~\bibnamefont {Bergues}}, \bibinfo {author} {\bibfnamefont {M.}~\bibnamefont {K{\"u}bel}}, \bibinfo {author} {\bibfnamefont {N.~G.}\ \bibnamefont {Johnson}}, \bibinfo {author} {\bibfnamefont {B.}~\bibnamefont {Fischer}}, \bibinfo {author} {\bibfnamefont {N.}~\bibnamefont {Camus}}, \bibinfo {author} {\bibfnamefont {K.~J.}\ \bibnamefont {Betsch}}, \bibinfo {author} {\bibfnamefont {O.}~\bibnamefont {Herrwerth}}, \bibinfo {author} {\bibfnamefont {A.}~\bibnamefont {Senftleben}}, \bibinfo {author} {\bibfnamefont {A.~M.}\ \bibnamefont {Sayler}}, \bibinfo {author} {\bibfnamefont {T.}~\bibnamefont {Rathje}}, \bibinfo {author} {\bibfnamefont {T.}~\bibnamefont {Pfeifer}}, \bibinfo {author} {\bibfnamefont {I.}~\bibnamefont {{Ben-Itzhak}}}, \bibinfo {author} {\bibfnamefont {R.~R.}\ \bibnamefont {Jones}}, \bibinfo {author} {\bibfnamefont {G.~G.}\ \bibnamefont {Paulus}}, \bibinfo {author} {\bibfnamefont {F.}~\bibnamefont {Krausz}}, \bibinfo {author} {\bibfnamefont
  {R.}~\bibnamefont {Moshammer}}, \bibinfo {author} {\bibfnamefont {J.}~\bibnamefont {Ullrich}},\ and\ \bibinfo {author} {\bibfnamefont {M.~F.}\ \bibnamefont {Kling}},\ }\bibfield  {title} {\bibinfo {title} {Attosecond tracing of correlated electron-emission in non-sequential double ionization},\ }\href {https://doi.org/10.1038/ncomms1807} {\bibfield  {journal} {\bibinfo  {journal} {Nat Commun}\ }\textbf {\bibinfo {volume} {3}},\ \bibinfo {pages} {813} (\bibinfo {year} {2012})}\BibitemShut {NoStop}%
\bibitem [{\citenamefont {Liu}\ \emph {et~al.}(2024)\citenamefont {Liu}, \citenamefont {Skruszewicz}, \citenamefont {Sp{\"a}the}, \citenamefont {Zhang}, \citenamefont {Hell}, \citenamefont {Ying}, \citenamefont {Paulus}, \citenamefont {Kiss}, \citenamefont {Murari}, \citenamefont {Khalil}, \citenamefont {Cormier}, \citenamefont {Jiao}, \citenamefont {Fritzsche},\ and\ \citenamefont {K{\"u}bel}}]{Liu2024_Exploring}%
  \BibitemOpen
  \bibfield  {author} {\bibinfo {author} {\bibfnamefont {F.}~\bibnamefont {Liu}}, \bibinfo {author} {\bibfnamefont {S.}~\bibnamefont {Skruszewicz}}, \bibinfo {author} {\bibfnamefont {J.}~\bibnamefont {Sp{\"a}the}}, \bibinfo {author} {\bibfnamefont {Y.}~\bibnamefont {Zhang}}, \bibinfo {author} {\bibfnamefont {S.}~\bibnamefont {Hell}}, \bibinfo {author} {\bibfnamefont {B.}~\bibnamefont {Ying}}, \bibinfo {author} {\bibfnamefont {G.~G.}\ \bibnamefont {Paulus}}, \bibinfo {author} {\bibfnamefont {B.}~\bibnamefont {Kiss}}, \bibinfo {author} {\bibfnamefont {K.}~\bibnamefont {Murari}}, \bibinfo {author} {\bibfnamefont {M.}~\bibnamefont {Khalil}}, \bibinfo {author} {\bibfnamefont {E.}~\bibnamefont {Cormier}}, \bibinfo {author} {\bibfnamefont {L.~G.}\ \bibnamefont {Jiao}}, \bibinfo {author} {\bibfnamefont {S.}~\bibnamefont {Fritzsche}},\ and\ \bibinfo {author} {\bibfnamefont {M.}~\bibnamefont {K{\"u}bel}},\ }\bibfield  {title} {\bibinfo {title} {Exploring valence-electron dynamics of xenon through laser-induced electron
  diffraction},\ }\href {https://doi.org/10.1103/PhysRevA.110.013118} {\bibfield  {journal} {\bibinfo  {journal} {Phys. Rev. A}\ }\textbf {\bibinfo {volume} {110}},\ \bibinfo {pages} {013118} (\bibinfo {year} {2024})}\BibitemShut {NoStop}%
\bibitem [{\citenamefont {Tsai}\ \emph {et~al.}(2022)\citenamefont {Tsai}, \citenamefont {Liang}, \citenamefont {Tsai}, \citenamefont {Lai}, \citenamefont {Lin},\ and\ \citenamefont {Chen}}]{Tsai2022_Nonlinear}%
  \BibitemOpen
  \bibfield  {author} {\bibinfo {author} {\bibfnamefont {M.-S.}\ \bibnamefont {Tsai}}, \bibinfo {author} {\bibfnamefont {A.-Y.}\ \bibnamefont {Liang}}, \bibinfo {author} {\bibfnamefont {C.-L.}\ \bibnamefont {Tsai}}, \bibinfo {author} {\bibfnamefont {P.-W.}\ \bibnamefont {Lai}}, \bibinfo {author} {\bibfnamefont {M.-W.}\ \bibnamefont {Lin}},\ and\ \bibinfo {author} {\bibfnamefont {M.-C.}\ \bibnamefont {Chen}},\ }\bibfield  {title} {\bibinfo {title} {Nonlinear compression toward high-energy single-cycle pulses by cascaded focus and compression},\ }\href {https://doi.org/10.1126/sciadv.abo1945} {\bibfield  {journal} {\bibinfo  {journal} {Sci. Adv.}\ }\textbf {\bibinfo {volume} {8}},\ \bibinfo {pages} {eabo1945} (\bibinfo {year} {2022})}\BibitemShut {NoStop}%
\bibitem [{\citenamefont {Brahms}\ \emph {et~al.}(2020)\citenamefont {Brahms}, \citenamefont {Belli},\ and\ \citenamefont {Travers}}]{Brahms2020_Infrared}%
  \BibitemOpen
  \bibfield  {author} {\bibinfo {author} {\bibfnamefont {C.}~\bibnamefont {Brahms}}, \bibinfo {author} {\bibfnamefont {F.}~\bibnamefont {Belli}},\ and\ \bibinfo {author} {\bibfnamefont {J.~C.}\ \bibnamefont {Travers}},\ }\bibfield  {title} {\bibinfo {title} {Infrared attosecond field transients and {{UV}} to {{IR}} few-femtosecond pulses generated by high-energy soliton self-compression},\ }\href {https://doi.org/10.1103/PhysRevResearch.2.043037} {\bibfield  {journal} {\bibinfo  {journal} {Phys. Rev. Res.}\ }\textbf {\bibinfo {volume} {2}},\ \bibinfo {pages} {043037} (\bibinfo {year} {2020})}\BibitemShut {NoStop}%
\bibitem [{\citenamefont {Jullien}\ \emph {et~al.}(2006)\citenamefont {Jullien}, \citenamefont {Kourtev}, \citenamefont {Albert}, \citenamefont {Cheriaux}, \citenamefont {Etchepare}, \citenamefont {Minkovski},\ and\ \citenamefont {Saltiel}}]{Jullien2006_Highly}%
  \BibitemOpen
  \bibfield  {author} {\bibinfo {author} {\bibfnamefont {A.}~\bibnamefont {Jullien}}, \bibinfo {author} {\bibfnamefont {S.}~\bibnamefont {Kourtev}}, \bibinfo {author} {\bibfnamefont {O.}~\bibnamefont {Albert}}, \bibinfo {author} {\bibfnamefont {G.}~\bibnamefont {Cheriaux}}, \bibinfo {author} {\bibfnamefont {J.}~\bibnamefont {Etchepare}}, \bibinfo {author} {\bibfnamefont {N.}~\bibnamefont {Minkovski}},\ and\ \bibinfo {author} {\bibfnamefont {S.}~\bibnamefont {Saltiel}},\ }\bibfield  {title} {\bibinfo {title} {Highly efficient temporal cleaner for femtosecond pulses based on cross-polarized wave generation in a dual crystal scheme},\ }\href {https://doi.org/10.1007/s00340-006-2334-7} {\bibfield  {journal} {\bibinfo  {journal} {Appl. Phys. B}\ }\textbf {\bibinfo {volume} {84}},\ \bibinfo {pages} {409} (\bibinfo {year} {2006})}\BibitemShut {NoStop}%
\bibitem [{\citenamefont {Smijesh}\ \emph {et~al.}(2019)\citenamefont {Smijesh}, \citenamefont {Zhang}, \citenamefont {Fischer}, \citenamefont {Muschet}, \citenamefont {Salh}, \citenamefont {Tajalli}, \citenamefont {Morgner},\ and\ \citenamefont {Veisz}}]{Smijesh2019_Contrast}%
  \BibitemOpen
  \bibfield  {author} {\bibinfo {author} {\bibfnamefont {N.}~\bibnamefont {Smijesh}}, \bibinfo {author} {\bibfnamefont {X.}~\bibnamefont {Zhang}}, \bibinfo {author} {\bibfnamefont {P.}~\bibnamefont {Fischer}}, \bibinfo {author} {\bibfnamefont {{\relax AA}.}~\bibnamefont {Muschet}}, \bibinfo {author} {\bibfnamefont {R.}~\bibnamefont {Salh}}, \bibinfo {author} {\bibfnamefont {A.}~\bibnamefont {Tajalli}}, \bibinfo {author} {\bibfnamefont {U.}~\bibnamefont {Morgner}},\ and\ \bibinfo {author} {\bibfnamefont {L.}~\bibnamefont {Veisz}},\ }\bibfield  {title} {\bibinfo {title} {Contrast improvement of sub-4 fs laser pulses using nonlinear elliptical polarization rotation},\ }\href {https://doi.org/10.1364/OL.44.004028} {\bibfield  {journal} {\bibinfo  {journal} {Opt. Lett.}\ }\textbf {\bibinfo {volume} {44}},\ \bibinfo {pages} {4028} (\bibinfo {year} {2019})}\BibitemShut {NoStop}%
\bibitem [{\citenamefont {Cerullo}\ \emph {et~al.}(2011)\citenamefont {Cerullo}, \citenamefont {Baltu{\v s}ka}, \citenamefont {M{\"u}cke},\ and\ \citenamefont {Vozzi}}]{Cerullo2011_Fewopticalcycle}%
  \BibitemOpen
  \bibfield  {author} {\bibinfo {author} {\bibfnamefont {G.}~\bibnamefont {Cerullo}}, \bibinfo {author} {\bibfnamefont {A.}~\bibnamefont {Baltu{\v s}ka}}, \bibinfo {author} {\bibfnamefont {O.}~\bibnamefont {M{\"u}cke}},\ and\ \bibinfo {author} {\bibfnamefont {C.}~\bibnamefont {Vozzi}},\ }\bibfield  {title} {\bibinfo {title} {Few-optical-cycle light pulses with passive carrier-envelope phase stabilization},\ }\href {https://doi.org/10.1002/lpor.201000013} {\bibfield  {journal} {\bibinfo  {journal} {Laser Photonics Rev.}\ }\textbf {\bibinfo {volume} {5}},\ \bibinfo {pages} {323} (\bibinfo {year} {2011})}\BibitemShut {NoStop}%
\bibitem [{\citenamefont {Moon}\ \emph {et~al.}(2006)\citenamefont {Moon}, \citenamefont {Li}, \citenamefont {Duan}, \citenamefont {Tackett}, \citenamefont {Corwin}, \citenamefont {Washburn},\ and\ \citenamefont {Chang}}]{Moon2006_Reductionc}%
  \BibitemOpen
  \bibfield  {author} {\bibinfo {author} {\bibfnamefont {E.}~\bibnamefont {Moon}}, \bibinfo {author} {\bibfnamefont {C.}~\bibnamefont {Li}}, \bibinfo {author} {\bibfnamefont {Z.}~\bibnamefont {Duan}}, \bibinfo {author} {\bibfnamefont {J.}~\bibnamefont {Tackett}}, \bibinfo {author} {\bibfnamefont {K.~L.}\ \bibnamefont {Corwin}}, \bibinfo {author} {\bibfnamefont {B.~R.}\ \bibnamefont {Washburn}},\ and\ \bibinfo {author} {\bibfnamefont {Z.}~\bibnamefont {Chang}},\ }\bibfield  {title} {\bibinfo {title} {Reduction of fast carrier-envelope phase jitter in femtosecond laser amplifiers},\ }\href {https://doi.org/10.1364/OE.14.009758} {\bibfield  {journal} {\bibinfo  {journal} {Opt. Express}\ }\textbf {\bibinfo {volume} {14}},\ \bibinfo {pages} {9758} (\bibinfo {year} {2006})}\BibitemShut {NoStop}%
\bibitem [{\citenamefont {Adachi}\ \emph {et~al.}(2004)\citenamefont {Adachi}, \citenamefont {Kumbhakar},\ and\ \citenamefont {Kobayashi}}]{Adachi2004_Quasimonocyclic}%
  \BibitemOpen
  \bibfield  {author} {\bibinfo {author} {\bibfnamefont {S.}~\bibnamefont {Adachi}}, \bibinfo {author} {\bibfnamefont {P.}~\bibnamefont {Kumbhakar}},\ and\ \bibinfo {author} {\bibfnamefont {T.}~\bibnamefont {Kobayashi}},\ }\bibfield  {title} {\bibinfo {title} {Quasi-monocyclic near-infrared pulses with a stabilized carrier-envelope phase characterized by noncollinear cross-correlation frequency-resolved optical gating},\ }\href {https://doi.org/10.1364/OL.29.001150} {\bibfield  {journal} {\bibinfo  {journal} {Opt. Lett., OL}\ }\textbf {\bibinfo {volume} {29}},\ \bibinfo {pages} {1150} (\bibinfo {year} {2004})}\BibitemShut {NoStop}%
\bibitem [{\citenamefont {Weckwerth}\ \emph {et~al.}(2025)\citenamefont {Weckwerth}, \citenamefont {Howard}, \citenamefont {Cheng}, \citenamefont {Gabalski}, \citenamefont {Ghrist}, \citenamefont {Mohideen}, \citenamefont {Lin}, \citenamefont {Yuen},\ and\ \citenamefont {Bucksbaum}}]{Weckwerth2025_Data}%
  \BibitemOpen
  \bibfield  {author} {\bibinfo {author} {\bibfnamefont {E.}~\bibnamefont {Weckwerth}}, \bibinfo {author} {\bibfnamefont {A.~J.}\ \bibnamefont {Howard}}, \bibinfo {author} {\bibfnamefont {C.}~\bibnamefont {Cheng}}, \bibinfo {author} {\bibfnamefont {I.}~\bibnamefont {Gabalski}}, \bibinfo {author} {\bibfnamefont {A.~M.}\ \bibnamefont {Ghrist}}, \bibinfo {author} {\bibfnamefont {S.~A.}\ \bibnamefont {Mohideen}}, \bibinfo {author} {\bibfnamefont {C.-D.}\ \bibnamefont {Lin}}, \bibinfo {author} {\bibfnamefont {C.-H.}\ \bibnamefont {Yuen}},\ and\ \bibinfo {author} {\bibfnamefont {P.~H.}\ \bibnamefont {Bucksbaum}},\ }\href {https://doi.org/10.5281/zenodo.17137158} {\bibinfo {title} {Data for {{Optical}} interference effect in strong-field electronic coherence spectroscopy}} (\bibinfo {year} {2025}),\ \bibinfo {note} {\url{https://doi.org/10.5281/zenodo.17137158}}\BibitemShut {NoStop}%
\end{thebibliography}
\end{document}